\documentclass[12pt,a4paper]{article}

\usepackage{amsfonts} % math fonts
\usepackage{mathtools} % math misc
\usepackage{amssymb} % math symbols
\usepackage{subfigure} % labelled subfigures
\usepackage{graphicx}
\usepackage{color}
\usepackage{cite} % Collected citations in ranges
\usepackage[normalem]{ulem}

\graphicspath{{Figs/}}

\newcommand{\abs}[1]{|#1|} % Absolute value, fixed |
\newcommand{\refEQ}[1]{Eq.\,\eqref{#1}} % reference to equation ``eq. (n)''
\newcommand{\refEQS}[1]{Eqs.\,\eqref{#1}} % reference to equations ``eq. (n)'' then use \eqref{etc}
\newcommand{\REFEQS}[1]{Eqs.\,\eqref{#1}} % reference to equations ``Eq. (n)'' then use \eqref{etc}
\newcommand{\nHH}{\mathrm{H}^0}
\newcommand{\nHR}{\mathrm{R}^0}
\newcommand{\nHI}{A}
\newcommand{\nh}{h}
\newcommand{\nH}{H}
\newcommand{\nA}{A}
\newcommand{\cH}{H^\pm}

\newcommand{\cHp}{H^+}
\newcommand{\gL}{\gamma_L}
\newcommand{\gR}{\gamma_R}
\newcommand{\cb}{c_\beta}
\renewcommand{\sb}{s_\beta}

\newcommand{\cab}{c_{\beta\alpha}}
\newcommand{\sab}{s_{\beta\alpha}}
\newcommand{\tb}{t_\beta}
\newcommand{\tbinv}{\tb^{-1}}
\newcommand{\tti}{\tb+\tbinv}
\newcommand{\ctti}{\cab(\tti)}
\newcommand{\CTTI}{{\cab^2(\tti)}^2}
\newcommand{\CKM}{V}
\newcommand{\V}[1]{{\CKM_{#1}^{\phantom{\ast}}}}
\newcommand{\Vc}[1]{{\CKM_{#1}^\ast}}
\newcommand{\PMNS}{U}
\newcommand{\U}[1]{{\PMNS_{#1}^{\phantom{\ast}}}}
\newcommand{\Uc}[1]{{\PMNS_{#1}^\ast}}
\newcommand{\MU}{D_u}
\newcommand{\MD}{D_d}
\newcommand{\NU}{{N_u^{\phantom{\dagger}}}}
\newcommand{\NUd}{{N_u^\dagger}}
\newcommand{\Nu}[1]{{N_u^{(#1)}}}
\newcommand{\ND}{{N_d^{\phantom{\dagger}}}}
\newcommand{\NDd}{{N_d^\dagger}}
\newcommand{\Nd}[1]{{N_d^{(#1)}}}

\newcommand{\Nl}[1]{{N_\ell^{(#1)}}}

\newcommand{\hmt}{\nh\to \mu\tau}

\newcommand{\hbq}{\nh\to bq}

\newcommand{\hbs}{\nh\to bs}

\newcommand{\hbd}{\nh\to bd}

\newcommand{\thq}{t\to\nh q}

\newcommand{\thc}{t\to\nh c}

\newcommand{\thu}{t\to\nh u}

\newcommand{\sigstr}[2]{\mu_{\rm #1}^{#2}}
\begin{document}

%%%%%%%%%%
\begin{titlepage}

% Preprint numbers
% \hfill\begin{minipage}[r]{0.3\textwidth}\begin{flushright}  CFTP-15-xxx\\    IFIC-15-xxx \end{flushright} \end{minipage}
\hfill\begin{minipage}[r]{0.3\textwidth}\begin{flushright} IFIC/15-62 \end{flushright} \end{minipage}

\begin{center}

\vspace{1.0cm}

{\large \bf {Flavour Changing Higgs Couplings in a Class of Two Higgs Doublet Models}}

\vspace{1.0cm}

F. J. Botella  $^{a,}$\footnote{\texttt{fbotella@uv.es}}, 
G. C. Branco  $^{a,b,}$\footnote{\texttt{gbranco@tecnico.ulisboa.pt}}, 
M. Nebot $^{c,}$\footnote{\texttt{nebot@cftp.ist.utl.pt}}
and M. N. Rebelo $^{c,}$\footnote{\texttt{rebelo@tecnico.ulisboa.pt}}

\vspace{1.0cm}

\emph{$^a$ Departament de F\' \i sica Te\`orica and IFIC,
Universitat de Val\`encia-CSIC, E-46100, Burjassot, Spain.} \\
\emph{$^b$ Departamento de F\'\i sica and Centro de F\' \i sica Te\' orica
de Part\' \i culas (CFTP),
Instituto Superior T\' ecnico (IST), U. de Lisboa (UL), Av. Rovisco Pais, P-1049-001 Lisboa,
Portugal.} \\
\emph{$^c$ Centro de F\' \i sica Te\' orica de Part\' \i culas (CFTP),
Instituto Superior T\' ecnico (IST), U. de Lisboa (UL), Av. Rovisco Pais, P-1049-001 Lisboa,
Portugal.}

\end{center}

\vspace{3cm}

%%%%%%%%%%
\begin{abstract}
We analyse various flavour changing processes like $t\to hu,hc$, $h\to \tau e,\tau\mu$ as well as hadronic decays $h\to bs,bd$, in the framework of a class of two Higgs doublet models where there are flavour changing neutral scalar currents at tree level. These models have the remarkable feature of having these flavour-violating couplings entirely determined by the CKM and PMNS matrices as well as $\tan\beta$. The flavour structure of these scalar currents results from a symmetry of the Lagrangian and therefore it is natural and stable under the renormalization group. We show that in some of the models the rates of the above flavour changing processes can reach the discovery level at the LHC at 13 TeV even taking into account the stringent bounds on low energy processes, in particular $\mu\to e\gamma$.
\end{abstract}
%%%%%%%%%%

\end{titlepage}
%%%%%%%%%%%%%%%%%%%%

\newpage

%%%%%%%%%%%%%%%%%%%%%%%%%%%%%%%%%%%%%%%%
\section{Introduction\label{SEC:INTRO}}
%%%%%%%%%%%%%%%%%%%%%%%%%%%%%%%%%%%%%%%%
The second run of the LHC, at a center of mass energy $\sqrt s=13$ TeV, will provide an important probe of flavour 
changing couplings of the recently discovered scalar boson $\nh$ \cite{Aad:2012tfa,Chatrchyan:2012xdj}.
These couplings can contribute to rare top decays like $\thq$ $(q=u,c)$ and may also lead 
to leptonic flavour changing decays such as $\nh \to \tau^\pm \ell^\mp$  
$(\ell = \mu, e)$, as well as hadronic flavour-changing decays 
like $\hbs$ and $\hbd$. These decays are strongly suppressed 
in the Standard Model (SM) since these couplings vanish at tree level. However, Higgs 
Flavour Violating Neutral Couplings (HFVNC) can arise in many extensions of the SM, including in 
Two Higgs Doublet Models (2HDM) 
\cite{Gunion:1989we,Branco:2011iw}. Any extension of the SM with HFVNC
has to comply with the strict experimental limits on processes mediated by flavour 
changing neutral currents as well as with the limits on CP violating transitions leading,
for example, to electric dipole moments of quarks and leptons \cite{Raidal:2008jk}.

In this paper, we investigate the allowed strength of HFVNC in the 
framework of a class of 2HDM, denoted BGL models, first proposed for the quark 
sector \cite{Branco:1996bq}, generalised in \cite{Botella:2009pq}
and then extended to the leptonic sector \cite{Botella:2011ne}.  
These models have the remarkable feature
of having HFVNC, but with their flavour structure entirely determined by the Cabibbo-Kobayashi-Maskawa (CKM) \cite{Cabibbo:1963yz,Kobayashi:1973fv} and the Pontecorvo-Maki-Nakagawa-Sakata (PMNS) \cite{Pontecorvo:1967fh,Maki:1962mu} matrices, denoted $\CKM$ and $\PMNS$ respectively. 
HFVNC have been widely addressed in the literature \cite{Hall:1981bc,Han:2000jz,Kanemura:2005hr,Carcamo:2006dp,Davidson:2010xv,Harnik:2012pb,Dery:2013aba,Bhattacharyya:2013rya,Bhattacharyya:2014nja,Celis:2014asa,Dery:2014kxa,Campos:2014zaa,Sierra:2014nqa,Lee:2014rba,Heeck:2014qea,Crivellin:2015mga,deLima:2015pqa,Dorsner:2015mja,Varzielas:2015iva,Das:2015zwa,Varzielas:2015joa,Bhattacherjee:2015sia,Mao:2015hwa,Altunkaynak:2015twa,He:2015rqa,Chiang:2015cba,Crivellin:2015hha,Altmannshofer:2015esa,Cheung:2015yga}. The distinctive feature of BGL models is the fact that many of the most dangerous HFVNC are naturally suppressed by combinations of small mixing 
matrix elements and/or light fermion masses. This is achieved through the introduction of a symmetry in the Lagrangian and therefore the 
suppression is entirely natural. Another salient feature of BGL models is that depending on the specific model 
of this class, HFVNC exist either in the up or in the down sector,
but not in both sectors simultaneously.  Analogous considerations apply to the leptonic sector. 
We will pay special attention to the evaluation of the maximum rate at which these processes can occur in this 
class of models, without violating the stringent bounds on processes like $\mu \to e \gamma$. In the general 2HDM, in the so-called Higgs basis \cite{Georgi:1978ri,Lavoura:1994fv,Botella:1994cs,Branco:1999fs},
there are three neutral scalars, which we denote $\nHH$, $\nHR$ and $\nHI$. The couplings of $\nHH$ to fermions are flavour diagonal in the fermion mass eigenstate basis. On the other hand, in the general 
2HDM the fields $\nHR$ and $\nHI$ have HFVNC with arbitrary flavour structure. The remarkable feature 
of BGL models is the fact that the flavour structure of HFVNC only depends on 
$\CKM$ and $\PMNS$. Furthermore, the neutral scalar mass eigenstates are linear combinations of 
$\nHH$ and $\nHR$ together with the pseudoscalar neutral field $\nHI$. In these models 
the imposed symmetry restricts the scalar potential in such a way that 
it cannot violate CP either explicitly or spontaneously, therefore once we perform 
a rotation through the angle $\beta$ that takes the fields from the basis
chosen by the symmetry to the Higgs basis, the 
charged fields and the pseudoscalar field $\nHI$ are already physical fields. As a result, the two
other neutral physical fields are related to $\nHH$ and $\nHR$ through a single angle rotation.
In a previous work \cite{Botella:2014ska}, 
we have performed a detailed analysis of the allowed mass ranges for the new scalars under the assumption that 
the discovered Higgs $\nh$ 
can be identified with $\nHH$. We have then shown that in some of the BGL models 
these masses can be in the range of a few hundred GeV and thus within reach of the LHC 13 TeV run. In this paper, we 
work in the general case where $\nh$ is a mixture of $\nHH$ and $\nHR$, controlled by an angle denoted $\beta-\alpha$. 
The strength of the HFVNC of $\nh$ depend crucially on $\tan\beta=v_2/v_1$, with $v_i$ the vacuum expectation values of the scalar doublets, and $\cos(\beta-\alpha)$.  
BGL models have many  features in common with several of  the implementations of the minimal flavour violation 
hypothesis (MFV) \cite{Buras:2000dm,D'Ambrosio:2002ex,Bobeth:2005ck,Dery:2013aba}.
However, BGL models have the unique 
feature of coming from a symmetry, and as a result they have a reduced number of free parameters.
This allows for definite predictions once constraints on these parameters are imposed, taking into  account 
the present experimental bounds.

The challenge is to answer the following question: in some of the BGL models, can one have regions,
in the $\tan\beta$ versus $\alpha-\beta$ plane, where 
the HFVNC of $\nh$ are such that the rare processes $\thq$, 
$\hmt$ can occur at a rate consistent with discovery at LHC-13 TeV?
 Of course, these regions have to be consistent with the stringent constraints
 on all Standard Model (SM) processes associated to the Higgs production and its
 subsequent decays in the channels $ZZ$, $WW$, $\gamma\gamma$, $b\bar b$ and $\tau\bar \tau$.
 Furthermore the constraints derived from low energy phenomenology have to be considered: both those obtained in \cite{Botella:2014ska}
 and those new due to the presence of $\nHH$-$\nHR$ mixing.
Processes such as $\hbs$ and $\hbd$ are probably not within reach
of the LHC but become important for the physics of the future Linear Collider. 
In this  paper, we also address the question  of what BGL  models could lead to
the observation of such decays by the future Linear Collider.

The paper is  organised as follows. In the next section we briefly review the main features of 
BGL models in order to set the notation. In Section \ref{SEC:top} we analyse top flavour changing decays of the type 
$\thq$ ($q=u,c$) in the framework of BGL models with HFVNC in the up sector. In Section \ref{SEC:HFC} we 
perform an analysis of flavour changing Higgs decays in BGL models. In particular, we consider neutrino models
 with HFVNC in the charged lepton sector giving rise to $\nh\to \ell \tau$ decays.
Up type models are also considered, giving rise to $\hbs$ flavour violating decays. In Section \ref{SEC:correlations}
we investigate the discovery regions and the existing correlations for these decays in the framework of BGL models. 
In section \ref{SEC:Conclusions} we present our conclusions. The paper contains two appendices,
in Appendix \ref{AP:HiggsLHC}, we explain how the relevant Higgs experimental data has been incorporated into the analysis, 
and in appendix \ref{AP:MEG}, we give details relative to the low energy flavour constraint $\mu\to e\gamma$.

\clearpage
 
%%%%%%%%%%%%%%%%%%%%%%%%%%%%%%%%%%%%%%%%
\section{Main Features of BGL Models\label{SEC:BGL}}
%%%%%%%%%%%%%%%%%%%%%%%%%%%%%%%%%%%%%%%%

Our work is done in the context of Two Higgs Doublet Models. The quark Yukawa interactions 
are denoted by:
\begin{equation}
\begin{split}
{\mathcal{L}}_{Y} =&
-\overline{Q_{L}^{0}}\,\big[\Gamma_{1}\,\Phi_{1}+\Gamma_{2}\,\Phi_{2}\big]\,d_{R}^{0}
-\overline{Q_{L}^{0}}\,\big[\Delta_{1}\,\tilde{\Phi}_{1}+\Delta_{2}\,\tilde{\Phi}_{2}\big]\,u_{R}^{0}
\\ &
-\overline{L_{L}^{0}}\,\big[\Pi_{1}\,\Phi_{1}+\Pi_{2}\,\Phi_{2}\big]\,l_{R}^{0}
-\overline{L_{L}^{0}}\,\big[\Sigma_{1}\,\tilde{\Phi}_{1}+\Sigma_{2}\,\tilde{\Phi}_{2}\big]\,\nu_{R}^{0}+\text{h.c.}, \label{YukawaDirac1}
\end{split}
\end{equation}
where $\Gamma_i$, $\Delta_i$  $ \Pi _{i}$ and $\Sigma_{i}$ are matrices in flavour space.
The requirement that $\Gamma_{i}$,  $\Delta_{i}$  lead to tree level Flavour Changing Neutral Couplings (FCNC)
with strength completely controlled by the Cabibbo - Kobayashi - Maskawa matrix $\CKM$, was achieved by 
Branco, Grimus and Lavoura (BGL)  \cite{Branco:1996bq} by means of a symmetry imposed on the 
quark and scalar sector of the Lagrangian of the form: 
\begin{equation}
Q_{Lj}^{0}\mapsto \exp {(i\tau)}\, Q_{Lj}^{0}\, ,\qquad
u_{Rj}^{0}\mapsto \exp {(i2\tau)}\,u_{Rj}^{0}\, ,\qquad \Phi
_{2}\mapsto \exp {(i\tau)}\,\Phi_{2}\, ,  \label{S symetry up quarks}
\end{equation}
where $\tau \neq 0, \pi$, with all other quark fields transforming 
trivially under the symmetry. The index $j$ can be fixed as either 1,
2 or 3. Alternatively the symmetry may be chosen as:
\begin{equation}
Q_{Lj}^{0}\mapsto \exp {(i\tau)}\, Q_{Lj}^{0}\, ,\qquad
d_{Rj}^{0}\mapsto \exp {(i2\tau)}\, d_{Rj}^{0}\, ,\quad 
\Phi_{2}\mapsto \exp {(-i\tau)}\, \Phi_{2}\, .  \label{S symetry down quarks}
\end{equation} 
The symmetry given by \refEQ{S symetry up quarks} leads to Higgs
FCNC in the down sector, whereas the symmetry specified by 
\refEQ{S symetry down  quarks} leads to Higgs FCNC in the up
sector. These two alternative choices of symmetry combined with the
three possible ways of fixing  the index $j$ give rise to six
different realizations of 2HDM with the flavour structure, in the
quark sector, controlled by the $\CKM$ matrix. We call up-type BGL models
those with HFVNC in the down sector, coming from the symmetry given by
\refEQ{S symetry up quarks} and we identify each one of the three implementations 
by u-type, c-type or t-type depending on the value of the index $j$, respectively 1, 2 or 3.
Likewise for the down-type models.
In the leptonic sector with Dirac neutrinos there is perfect analogy with the quark sector
and the corresponding  symmetry applied to the leptonic fields leads to six different
realizations with the strength of Higgs mediated flavour changing neutral currents
controlled by the Pontecorvo - Maki - Nakagawa - Sakata matrix, $\PMNS$.
As a result there are thirty six different implementations of BGL models. 
As was shown in reference \cite{Botella:2011ne}, in the case of Majorana neutrino there are only 18 models
 corresponding to the neutrino types and therefore with HFVNC in the charged lepton sector.

The discrete symmetry of \REFEQS{S symetry up quarks} or \eqref{S symetry down quarks} constrains the Higgs potential to be of the form:
\begin{equation}
\begin{split}
V=&\ \mu_1 \Phi_1^{\dagger}\Phi_1+\mu_2\Phi_2^{\dagger}\Phi_2-m_{12}\left( \Phi_1^{\dagger}\Phi_2+ \Phi_2^{\dagger}\Phi_1\right)+
2\lambda_3\left(\Phi^{\dagger}_1\Phi_1\right)\left(\Phi_2^{\dagger}\Phi_2\right)\\
&+2\lambda_4\left(\Phi_1^{\dagger}\Phi_2\right)\left(\Phi_2^{\dagger}\Phi_1\right)+
\lambda_1\left(\Phi_1^{\dagger}\Phi_1\right)^2+
\lambda_2\left(\Phi_2^{\dagger}\Phi_2\right)^2,
\end{split}\label{eq:ScalarPotential}
\end{equation}
the term in $m_{12}$ is a soft symmetry breaking term. Its
introduction prevents the appearance of an would-be Goldstone boson
due to an accidental continuous global symmetry of the  potential,
which arises when the BGL symmetry is exact. Such a potential cannot violate CP
either explicitly or spontaneously. As a result the scalar and pseudoscalar 
neutral Higgs fields do not mix among themselves and we are left with only two important 
rotation angles, $\beta$ and $\alpha$. The angle $\beta$ is such that the orthogonal 
matrix parametrised  by this angle leads to the Higgs basis, singling out the three 
neutral fields:  $\nHH$, with couplings to the quarks proportional to mass matrices, 
$\nHR$ which is a scalar  neutral Higgs and $\nHI$ which is a pseudoscalar neutral Higgs; as 
well as the physical  charged  Higgs fields $\cH$ and the pseudo-Goldstone bosons.  
In BGL models the fields $\nHI$ and  $\cH$ are already physical Higgs fields, while
$\nHH$ and $\nHR$  may still mix. In the limit  in which  $\nHH$ does not mix with 
$\nHR$,  $\nHH$  is identified with the Higgs field $\nh$ recently discovered by  
ATLAS \cite{Aad:2012tfa}
and CMS \cite{Chatrchyan:2012xdj}. In this limit this field does not mediate tree level flavour changes and  $\alpha$
is defined in such a way that the mixing angle between these fields, $(\beta - \alpha) $,
acquires the value $\pi / 2$. In fact
expanding the neutral scalar fields around their vacuum expectation values \cite{Lee:1973iz}
in the form
$\phi^0_j=\frac{1}{\sqrt{2}}(v_j+\rho_j+i\eta_j)$ we can write:
\begin{equation}
\begin{pmatrix} \nHH \\ \nHR \end{pmatrix} \equiv
\begin{pmatrix}\cos \beta  & \sin \beta \\ -\sin \beta & \cos \beta \end{pmatrix}
\begin{pmatrix}\rho_1\\ \rho_2 \end{pmatrix}\,,\ 
\begin{pmatrix} \nH \\ \nh \end{pmatrix} \equiv	
\begin{pmatrix}\cos \alpha  & \sin \alpha \\ -\sin \alpha & \cos \alpha \end{pmatrix}
\begin{pmatrix}\rho_1\\ \rho_2\end{pmatrix}\ .
\label{beta}
\end{equation}
The angle $\beta$ is determined by $\tan \beta \equiv v_2 /v_1$; in the following we use the shorthand notation $\tan\beta\equiv\tb$, $\cos(\beta-\alpha)\equiv\cab$ and $\sin(\beta-\alpha)\equiv\sab$.

The general form for the Yukawa couplings of 2HDM written in terms of quark mass 
eigenstates and the scalar fields in the Higgs basis is given by:
\begin{equation}
\begin{split}
{\mathcal L}_{\rm Yuk} = & 
- \frac{\sqrt{2}\,\cHp}{v}\, \bar{u} \left[\CKM \ND \gR - \NUd \CKM \gL \right] d +  \text{h.c.}\\
& - \frac{\nHH}{v}\left[\bar{u}\, \MU\, u + \bar{d}\, \MD\, d \right]\\
& - \frac{\nHR}{v} \left[\bar{u}\,(\NU \gR + \NUd \gL)\,u+\bar{d}\,(\ND \gR + \NDd \gL)\, d \right] \\
& + i\frac{\nHI}{v}\left[\bar{u}\,(\NU \gR - \NUd \gL)\,u-\bar{d}\,(\ND \gR - \NDd \gL)\, d \right]
\label{eq6}
\end{split}
\end{equation}
where $\gL$ and $\gR$ are the left-handed and right-handed chirality projectors, 
respectively, and $\MD$ and $\MU$ are the diagonal mass matrices for down and up quarks respectively.
 This equation defines the matrices $\ND$ and $\NU$  which 
give the strength and the flavour structure of FCNC and  are also involved in the couplings 
of the charged Higgs fields. In general 2HDM, still in the Higgs basis, the flavour structure  
of the quark sector is fully  determined by the quark masses, the $\CKM$ matrix 
and the two matrices  $\ND$ and $\NU$.  It is worth emphasising the high predictive 
power of the general 2HDM, as can be seen from \refEQ{eq6}. Let us assume that a pair of charged 
Higgs $\cH$ and the three neutral scalars (in general the physical neutral scalars
are combinations of $\nHH$, $\nHR$ and $\nHI$) are discovered. The couplings of  $\cH$ 
to quarks can be readily measured from their decays. Since $\CKM$ in \refEQ{eq6}
stands for the CKM matrix which is known, from $\cH$ decays one can derive 
$\ND$ and $\NU$, which enables one to predict in the framework of the general
2HDM the flavour structure of the neutral scalar couplings. This would be essential to prove
that the discovered neutral and charged scalar particles were part of a two Higgs doublet 
structure. In general 2HDM, the matrices $\ND$ and $\NU$, are entirely arbitrary. On the 
contrary, BGL models have the remarkable feature of having the flavour structure of  
$\ND$ and $\NU$ entirely determined by fermion masses, $\CKM$  and the angle $\beta$
with no other free parameters. This results from the symmetry introduced in the Lagrangian,
in order to achieve the BGL flavour structure in a natural way. As previously emphasized,
each one of the six implementations of BGL in the quark sector only has FCNC in one of the quark 
sectors, either up or down. In BGL up-type models  the matrices $\ND$ and $\NU$ 
have the following simple form:
\begin{equation}
{\big(\Nd{u_j}\big)}_{rs} = \left[ \tb \delta_{rs} - (\tti) \Vc{jr}\V{js} \right]  {(\MD)}_{ss}\,,
 \label{24}
\end{equation}
where no sum in $j$ implied. The upper index  $(u_j)$ indicates that 
we are considering a symmetry  of the form given by \refEQ{S symetry up quarks}, 
i.e., an up-type model with index $j$, thus leading to
FCNC in the down-sector. Notice that all FCNC are proportional to the factor $(\tti) $
multiplying products of  entries involving one single row of $\CKM$. The corresponding $\NU$
matrix is given by:
\begin{equation}
{\big(\Nu{u_j}\big)}_{rs} = \left[ \tb - (\tti)\delta_{rj} \right] {(\MU)}_{ss}\, \delta_{rs}\,.
\label{25} 
\end{equation}
$\NU$ is a diagonal matrix, the $\tb$ dependence is not the same for each diagonal entry. 
It is  proportional to $-\tbinv$  for the $(jj)$ element and to $\tb$ for all other elements.  
The index $j$ fixes the row of  the $\CKM$ matrix which suppresses the flavour 
changing neutral currents. Since, for each up-type  BGL model a single row of  $\CKM$
participates in these couplings, one may choose a phase convention
where all elements of $\ND$ and $\NU$ are real. 
For down-type models, which  correspond to the symmetry given by 
\refEQ{S symetry down  quarks}, 
the matrices $\ND$ and $\NU$  exchange r\^ ole, and now we have:
\begin{equation}
{\big(\Nu{d_j}\big)}_{rs} =  \left[ \tb \delta_{rs} - (\tti) \V{rj}\Vc{sj} \right]  {(\MU)}_{ss}\,,
\label{ees}
\end{equation}
\begin{equation}
{\big(\Nd{d_j}\big)}_{rs} =  \left[ \tb  - (\tti)\delta_{rj} \right] {(\MD)}_{ss}\, \delta_{rs}\,.
\label{EQ:Nd:dj:01}
\end{equation}
In down-type models the flavour changing neutral currents are suppressed 
by the columns of the $\CKM$ matrix.

In this paper we allow for the possibility of $\nh$ being a linear combination of $\nHH$ and $\nHR$
which is parametrised by the angle $(\beta-\alpha)$:
\begin{equation}
\nh =  \sab\, \nHH + \cab\, \nHR\,, \quad
\nH =  \cab\, \nHH - \sab\, \nHR\,.
\label{EQ:Smix:01}
\end{equation}
This mixing is constrained by data from the LHC Higgs observables (see appendix \ref{AP:HiggsLHC}).
 The quark Yukawa couplings of the
Higgs field $\nh$  can be denoted as:
\begin{equation}
{\mathcal L}_{\nh q\bar q} = -Y^{D}_{ij} \,  \bar d_{Li}\, d_{Rj}\, \nh - Y^{U}_{ij}\, \bar u_{Li}\, u_{Rj}\, \nh + \text{h.c.}\,,\label{EQ:QuarkYukawas:01}
\end{equation}
and similarly for the leptonic sector with the coefficients denoted by
$Y^\ell_{ij}$ and $Y^\nu_{ij}$.
From \REFEQS{eq6} and \eqref{EQ:Smix:01}, we  can read:
\begin{equation}
\begin{split}
Y^{D}_{ij} &=  \frac{1}{v} \left[  \sab\, (\MD)_{ij} + \cab\, (\ND)_{ij} \right] \\
Y^{U}_{ij} &=  \frac{1}{v} \left[  \sab\, (\MU)_{ij} + \cab\, (\NU)_{ij} \right] \label{yuij}
\end{split}
\end{equation}
More specifically, for $i \neq j$, we get the following flavour violating Yukawa
couplings, for the different types of BGL models:
\begin{itemize}
\item[(i)] up-type $u_k$ model, with $k$ fixed as 1 ($u$) or 2 ($c$) or 3 ($t$), where 
HFVNC arise in the down quark sector:
\begin{equation}
Y^{D}_{ij}({u_k}) = -\Vc{ki} \V{kj}\,\frac{m_{d_j}}{v}\,\ctti\,,\ i\neq j,\text{ no sum in } k\,,\label{EQ:BGLup:DownYukawa:01}
\end{equation}
\item[(ii)] down-type $d_k$ model, with $k$ fixed as 1 ($d$) or 2 ($s$) or 3 ($b$), where 
HFVNC arise in the up quark sector:
\begin{equation}
Y^{U}_{ij} ({d_k})= -\V{ik} \Vc{jk}\,\frac{m_{u_j}}{v}\,\ctti\,,\ i\neq j,\text{ no sum in } k\,,\label{EQ:BGLdown:UpYukawa:01}
\end{equation}
\item[(iii)] leptonic sector, neutrino-type, $\nu_k$ model, with $k$ fixed as 1 ($\nu_1$) or 2 ($\nu_2$) or 3 ($\nu_3$), where 
HFVNC arise in the charged lepton sector:
\begin{equation}
Y^{\ell}_{ij}({\nu_k}) = -\U{ik} \Uc{jk}\,\frac{m_{l_j}}{v}\,\ctti\,,\ i\neq j,\text{ no sum in } k\,.
\label{bbb}
\end{equation}
\end{itemize}
In the case of Dirac neutrinos one can write similar expressions
for charged lepton type models and in this case the FCNC appear in the 
neutrino sector and are suppressed by the extremely small neutrino masses.
In the case of models of the charged lepton type, only diagonal couplings are relevant. These couplings, as all other diagonal ones, can be extracted from equations \eqref{24} -- \eqref{EQ:Nd:dj:01}, and if necessary making the corresponding changes of quarks by leptons.

\clearpage
 
%%%%%%%%%%%%%%%%%%%%%%%%%%%%%%%%%%%%%%%%
\section{Flavour changing decays of top quarks\label{SEC:top}}
%%%%%%%%%%%%%%%%%%%%%%%%%%%%%%%%%%%%%%%%

In this section, we analyse flavour changing decays of top quarks $\thq$.
 They can arise in down-type BGL models, where there are 
Higgs flavour violating neutral currents in the up sector. According to \refEQS{yuij} and \eqref{EQ:BGLdown:UpYukawa:01},
 the couplings of the Higgs particle $\nh$ with a top $t$ and a light up-type quark $u$ or $c$, in a model of type $d_\rho$, can be written as
\begin{equation}
Y^{U}_{qt}(d_\rho)= -\V{q\rho} \Vc{t\rho}\,\frac{m_{t}}{v}\,\ctti\,,\quad q=u,c\,.
\label{ltqh}
\end{equation}
One can then evaluate the corresponding $\thq$ decay rate. As previously mentioned, there
are three types of models of this class, $d_\rho$, depending on the column of the $\CKM$ matrix
 which suppresses the flavour changing currents. The result is
\begin{equation}
\Gamma_{(d_\rho)}(\thq) =\frac{m_{t}^{3}}{32\pi v^2}
\left( 1-\frac{m_{\nh}^{2}}{m_{t}^{2}}\right)^{2}
\abs{\V{q\rho}}^{2}\abs{\V{t\rho}}^{2}\CTTI\,.  \label{qqq}
\end{equation}
Note that, apart from the global factor $\CTTI$,
every other factor in \refEQ{qqq} is fixed once we choose the specific down-type model $d_\rho$
 and the decay channel $\thc$ or $\thu$. Therefore, for a given model, $\thq$ processes constrain the factor $\CTTI$. 
In Table \ref{TAB:thq:CKM} we enumerate the decay channels and the models according to the
$\CKM$ factors involved.
\begin{table}[hbt]
\begin{center}
\begin{tabular}{|c||c|c|}
\hline
Model& $\thu$ & $\thc$ \\ \hline
$d$ & $\abs{\V{ud}\V{td}}^{2}\,(\sim\lambda^{6})=7.51\cdot 10^{-5}$ & $\abs{\V{cd}\V{td}}^{2}\,(\sim\lambda^{8})=4.01\cdot 10^{-6}$ \\ \hline
$s$ & $\abs{\V{us}\V{ts}}^{2}\,(\sim\lambda^{6})=8.20\cdot 10^{-5}$ & $\abs{\V{cs}\V{ts}}^{2}\,(\sim\lambda^{4})=1.53\cdot 10^{-3}$ \\ \hline
$b$ & $\abs{\V{ub}\V{tb}}^{2}\,(\sim\lambda^{6})=1.40\cdot 10^{-5}$ & $\abs{\V{cb}\V{tb}}^{2}\,(\sim\lambda^{4})=1.68\cdot 10^{-3}$ \\ \hline
\end{tabular}
\caption{$\CKM$ factors entering \refEQ{qqq}, $\lambda\simeq 0.22$ \cite{Agashe:2014kda} is the Cabibbo angle \cite{Cabibbo:1963yz} or Wolfenstein main expansion parameter \cite{Wolfenstein:1983yz}.\label{TAB:thq:CKM}}
\end{center}
\end{table}

It is clear that the most interesting models for $\thc$ are the $s$ and $b$ models,
 where the suppression is only at the $\lambda^{4}$ level, compared to the $d$ model
 which has a strong suppression for the same decay at the $\lambda^{8}$ level.
The $d$ model has the curiosity that the suppression is higher for $\thc$ than for $\thu$, unlike in the other two models. 
The branching ratio for $\thq$ in the $d_{\rho}$ type model is
\begin{equation}
\text{Br}_{(d_\rho)}(\thq)=
\frac{\Gamma_{(d_\rho)}(\thq)}{\Gamma(t\to Wb)}
=f(x_\nh,y_W)\,\frac{\abs{\V{q\rho}\V{t\rho}}^2}{\abs{\V{tb}}^2}\,\CTTI\,,
\label{EQ:BRthq:01}
\end{equation}
where
\begin{equation}
f(x_\nh,y_W)=\frac{1}{2}\left(1-x_\nh\right)^{2}\left(1-3y_W^2+2y_W^{3}\right)^{-1}\,,\text{ with } x_\nh=\frac{m_\nh^2}{m_t^2}\,,\ y_W=\frac{M_W^2}{m_t^2}\,.
\end{equation}
Using the top quark pole mass $m_{t}=173.3$ GeV \cite{Agashe:2014kda}, $m_{\nh}=125.0$ GeV and $M_{W}=80.385$ GeV, one obtains $f(x_\nh,y_W)=0.1306$\,.
Considering then the upper bounds $0.79\%$ from the ATLAS \cite{Aad:2014dya} and $0.56\%$ from the CMS \cite{Chatrchyan:2014aea,Khachatryan:2014jya} collaborations, we obtain, for $b$ and $s$-type models, the following constraint
\begin{equation}
\abs{\ctti} \lesssim 4.9\,.\label{eq:BRthq:ctti:01}
\end{equation}
Notice that for this value, perturbative unitarity constraints have to be considered (see appendix \ref{AP:Pert}).

% \clearpage
 
%%%%%%%%%%%%%%%%%%%%%%%%%%%%%%%%%%%%%%%%
\section{Flavour changing Higgs decays\label{SEC:HFC}}
%%%%%%%%%%%%%%%%%%%%%%%%%%%%%%%%%%%%%%%%
%%%%%%%%%%%%%%%%%%%%%%%%%
\subsection{The decays  $\boldsymbol{h\to \ell \tau}$  ($\boldsymbol{\ell =  \mu, e}$)}
%%%%%%%%%%%%%%%%%%%%%%%%%
The most interesting BGL models with HFVNC in the leptonic sector are the $\nu$ models,
where there are FCNC in the charged lepton sector. As seen in the previous section
for the quark sector, there are three neutrino-type BGL models, depending on the column
 of the $\PMNS$ matrix which enters the FCNC in the leptonic sector. Using a notation analogous to the one
of the quark sector and considering \refEQ{bbb} for the $\nh$ coupling to $\mu$ and $\tau$, we have
\begin{equation}
Y^{\ell}_{\mu \tau}(\nu_\rho)=
\frac{1}{v}\,\cab\,\big(\Nl{\nu_{\sigma}}\big)_{\mu\tau}=
 -\ctti\, \U{\mu\sigma}\Uc{\tau\sigma}\, \frac{m_{\tau}}{v}\,,
\end{equation}
and the decay rate:
\begin{equation}
\Gamma_{(\nu_\sigma)}(\nh\to \mu \bar\tau)
+
\Gamma_{(\nu_\sigma)}(\nh\to \bar\mu \tau)
=\CTTI\abs{\U{\mu\sigma}\U{\tau\sigma}}^{2}\,\Gamma_{\rm SM}(\nh\to\tau\bar\tau)\,,
\label{EQ:hmt:Gamma:01}
\end{equation}
with $\Gamma_{\rm SM}(\nh\to\tau\bar\tau)= \frac{m_\nh}{8\pi}\frac{m_\tau^2}{v^2}$.
Notice, again, the appearance of the same factor $\CTTI$. Table \ref{TAB:hmt:PMNS} lists the PMNS mixing matrix factors for the 
different $\nu$ - type models.
\begin{table}[h]
\begin{center}
\begin{tabular}{|c||c|c|c|}
\hline
Model & $\nh\to e\mu$ & $\nh\to e\tau$ & $\nh\to \mu\tau$ \\ \hline
$\nu_{1}$ & $\abs{\U{e1}\U{\mu 1}}^{2}(\sim \frac{1}{9})=0.105$ & $\abs{\U{e1}\U{\tau 1}}^{2}(\sim \frac{1}{9})=0.118$ & $\abs{\U{\mu 1}\U{\tau 1}}^{2}(\sim \frac{1}{36})=0.028$ \\ \hline
$\nu_{2}$ & $\abs{\U{e2}\U{\mu 2}}^{2}(\sim \frac{1}{9})=0.089$ & $\abs{\U{e2}\U{\tau 2}}^{2}(\sim \frac{1}{9})=0.126$ & $\abs{\U{\mu 2}\U{\tau 2}}^{2}(\sim \frac{1}{9})=0.115$ \\ \hline
$\nu_{3}$ & $\abs{\U{e3}\U{\mu 3}}^{2}=0.0128$ & $\abs{\U{e3}\U{\tau 3}}^{2}=0.0097$ & $\abs{\U{\mu 3}\U{\tau 3}}^{2}(\sim \frac{1}{4})=0.234$ \\ \hline
\end{tabular}
\caption{$\PMNS$ factors entering \refEQ{EQ:hmt:Gamma:01} for the  different $\nu$ - type models; estimates, e.g. $1/9$, $1/36$, correspond to a tri-bimaximal $\PMNS$ (except, of course, for $\abs{\PMNS_{e3}}$); analogous information for $\nh\to e\mu$ and $\nh\to e\tau$ decays is provided.\label{TAB:hmt:PMNS}}
\end{center}
\end{table}

The first direct search for lepton-flavour-violating decays of the observed Higgs boson 
performed by the CMS collaboration \cite{Khachatryan:2015kon}, led to the observation of a slight 
excess of signal events with a significance of 2.4 standard deviations.
 The best fit value is:
\begin{equation}
\text{Br}(\nh\to \mu\bar\tau+\tau\bar\mu)=\left( 0.84\begin{smallmatrix}+0.39\\ -0.37\end{smallmatrix}\right)\%\,,
\label{bla}
\end{equation}
which sets a constraint on the branching fraction $\text{Br}(\nh\to \mu\bar\tau+\tau\bar\mu) < 1.51 \% $ at the 95\%  confidence level.
 The ATLAS collaboration has presented a result based on hadronic $\tau$ decays \cite{Aad:2015gha}, giving $\text{Br}(\nh\to \mu\bar\tau+\tau\bar\mu) =(0.77\pm 0.62) \% $.
Assuming the $\nh$ width to be $\Gamma_\nh\simeq\Gamma_\nh^{[\rm SM]}(=4.03\text{ MeV})$, one can use the SM branching ratio $\text{Br}_{\rm SM}(\nh\to \tau\bar\tau)=0.0637$ in \refEQ{EQ:hmt:Gamma:01}, and obtain the estimate
\begin{equation}
\abs{\ctti}\sim 1\,,\label{eq:ctti_hmt:00}
\end{equation}
necessary to produce $\text{Br}(\nh\to \mu\bar\tau+\tau\bar\mu)$ of order $10^{-2}$.
%%%%%%%%%%%%%%%%%%%%%%%%%
\subsection{The flavour changing decays  $\boldsymbol{\hbq}$  ($\boldsymbol{q =  s, d}$)}
%%%%%%%%%%%%%%%%%%%%%%%%%

We now address up-type BGL models, where there are scalar mediated 
FCNC in the down sector and the most promising experimental signatures correspond to $\hbq$ decays, with $q=s,d$.
 The relevant flavour changing $\nh$ couplings to the down quarks in \refEQ{EQ:QuarkYukawas:01} are, according to \refEQ{EQ:BGLup:DownYukawa:01}:
\begin{equation}
Y^{D}_{qb}({u_k}) = -\ctti\,\Vc{kq} \V{kb}\,\frac{m_{b}}{v}\,,\ q\neq b,\text{ no sum in } k\,.
\end{equation}
Once again, it should be emphasised that once the up-type model $u_k$ is chosen, the
strength of the flavour changing couplings only depends on the combination $\ctti$ together with the down quark masses and $\CKM$ factors which are already known. The decay rate of $\nh$ to pairs of quarks $q_iq_j$ ($i\neq j$) is
\begin{equation}
\Gamma_{(u_k)}(\nh\to \bar q_iq_j+q_i\bar q_j)=\frac{3\,m_\nh}{8\pi}\left[\frac{1}{2}\abs{Y_{ij}}^2 + \frac{1}{2}\abs{Y_{ji}}^2 \right]\,.
\end{equation}
We thus have
\begin{equation}
\Gamma_{(u_k)}(\nh\to \bar bq+b\bar q)=\CTTI\,\abs{\V{kq}}^2\abs{\V{kb}}^2\,\Gamma_{\rm SM}(\nh\to b\bar b)\,.\label{EQ:Gamma:hbq:01}
\end{equation}
Assuming that $\Gamma_\nh\simeq \Gamma_\nh^{[\rm SM]}$, we can make the following estimate
\begin{equation}
\text{Br}_{(u_k)}(\nh\to \bar bq+b\bar q)=\CTTI\,\abs{\V{kq}}^2\abs{\V{kb}}^2\,\text{Br}_{\rm SM}(\nh\to b\bar b)\,,\label{EQ:Gamma:hbq:02}
\end{equation}
where $\text{Br}_{\rm SM}(\nh\to b\bar b)=0.578$. The relevant CKM-related factors for $\hbs$ and $\hbd$ in all three $u_k$ BGL models are given in Table \ref{TAB:hbq:CKM}. 
\begin{table}[hbt]
\begin{center}
\begin{tabular}{|c||c|c|}
\hline
Model & $\hbd$ & $\hbs$ \\ \hline
$u$ & $\abs{\V{ud}\V{ub}}^{2}\,(\sim \lambda^6)=1.33\cdot 10^{-5}$ & $\abs{\V{us}\V{ub}}^{2}\,(\sim \lambda^8)=7.14\cdot 10^{-7}$ \\ \hline
$c$ & $\abs{\V{cd}\V{cb}}^{2}\,(\sim \lambda^6)=8.52\cdot 10^{-5}$ & $\abs{\V{cs}\V{cb}}^{2}\,(\sim \lambda^4)=1.59\cdot 10^{-3}$ \\ \hline
$t$ & $\abs{\V{td}\V{tb}}^{2}\,(\sim \lambda^6)=7.90\cdot 10^{-5}$ & $\abs{\V{ts}\V{tb}}^{2}\,(\sim \lambda^4)=1.61\cdot 10^{-3}$ \\ \hline
\end{tabular}
\caption{$\CKM$ factors entering \refEQ{EQ:Gamma:hbq:01}, $\lambda\simeq 0.22$.\label{TAB:hbq:CKM}}
\end{center}
\end{table}

We thus have, to a good approximation:
\begin{itemize}
\item in models $c$ and $t$,
\begin{equation}
\text{Br}(\nh\to \bar bs+b\bar s)\sim \CTTI\,\lambda^4\,\sim 10^{-3}\,\CTTI\,,
\end{equation}
\item in model $u$,
\begin{equation}
\text{Br}(\nh\to \bar bs+b\bar s)\sim \CTTI\,\lambda^8\,\sim 10^{-7}\,\CTTI\,,
\end{equation}
\item in all $u$, $c$ and $t$ models,
\begin{equation}
\text{Br}(\nh\to \bar bd+b\bar d)\sim \CTTI\,\lambda^6\,\sim 10^{-5}\,\CTTI\,.
\end{equation}
\end{itemize}
We stress that, a priori, in models where there is no $\hmt$ constraint, one can reach values for $\text{Br}(\nh\to b\bar s+s\bar b)$ not far from $10^{-1}$. This can happen in charged lepton models of the charm and top types with $\ctti$ ranging from 5 to 10. 
Again, see appendix \ref{AP:Pert} for perturbative unitarity constraints on these values of $\ctti$.

% \clearpage
 
%%%%%%%%%%%%%%%%%%%%%%%%%%%%%%%%%%%%%%%%
\section{Correlations among Observables\label{SEC:correlations}}
%%%%%%%%%%%%%%%%%%%%%%%%%%%%%%%%%%%%%%%%
One of the most interesting aspects of flavour violation in BGL models is the 
fact that, in this framework, it is possible to establish clear correlations between 
various flavour violating processes. As previously emphasized, one of the key features
of the BGL models analysed in this paper is the presence of flavour 
changing neutral currents at tree level, but naturally suppressed by entries of
the CKM and/or PMNS mixing matrices. Apart from these mixing 
matrices, in these models FCNC only depend on the values of  $\tan\beta$, 
$\cos({\beta - \alpha})$ and fermion masses.
However, the level of suppression depends on the specific BGL model, which in turn implies that the correlations 
differ from model to model. 
In this paper we analyse the tree level flavour violating decays involving the 
Higgs boson already discovered at the LHC,
which were listed in the previous section. It should be pointed
out that the analysis has to take into consideration the flavour conserving
Higgs constraints already obtained from Run 1 of the LHC. In particular one has to
comply with the measured signal strengths for the five relevant decay channels
measured, to wit: $\nh \to W^+ W^-$,
$\nh \to ZZ$, $\nh \to b \bar b$, $\nh \to \tau\bar\tau$,
and $\nh \to \gamma \gamma$. They involve the flavour
conserving couplings of the Higgs and put constraints in the $\tb$ vs. $\alpha-\beta$ available space.
 A detailed explanation of observables and the constraints is given in appendix \ref{AP:HiggsLHC}.
 An extended analysis of the phenomenology of the models under consideration,
 of the type presented in \cite{Botella:2014ska}, is beyond the scope of this paper.
 However we also take into consideration
the current bounds from the low energy processes $B_d - \bar B_d$ mixing, 
$B_s - \bar B_s$ mixing, $K^0 - \bar K^0$ mixing and $D^0 - \bar D^0$  mixing.   
The experimental limits on  these processes can be, in principle, easily translated into limits
on the combination $\CTTI$ which is relevant for our models,
 as will be shown in what follows.
 As pointed 
out before, there are several different types of BGL models and depending on the model 
under consideration we have FCNC in the down sector or in the up sector, but 
never in both sectors at the same time. We can thus analyse different kinds of correlations among the different HFVNC observables considered in the previous sections:
\begin{itemize}
\item within the same quark sector:
\begin{itemize}
\item $\thc$ vs. $\thu$, in down-type models, where there are tree level FCNC in the up quark sector,
\item $\hbs$ vs. $\hbd$, in up-type models, where there are tree level FCNC in the down quark sector,
\end{itemize}
\item within the quark and the lepton sector in neutrino-type models (where there are tree level FCNC in the charged lepton sector),
\begin{itemize}
\item $\thq$ vs. $\hmt$, in down-neutrino-type models, 
\item $\hbq$ vs. $\hmt$, in up-neutrino-type models.
\end{itemize}
\end{itemize}
Taking into account the different features of specific BGL models,
we will present various correlations in the framework of BGL models where the
above processes may occur at a level consistent with discovery
at the LHC and/or at the future Linear Collider.
Notice that, at present, there is evidence \cite{Khachatryan:2015kon} from the CMS collaboration -- not challenged by the recent ATLAS result \cite{Aad:2015gha} -- pointing towards the possible observation of the decay $\hmt$, which would constitute a clear signal of physics beyond 
the SM. If such an evidence persists it will favour neutrino-type BGL models. Top quark decays such as $\thc$ and $\thu$
 will be further explored in Run 2 of the LHC. Concerning $\hbs$ and $\hbd$, these decays are probably inaccessible at the LHC due to the overwhelming backgrounds. However, they constitute very promising channels for the future Linear Collider \cite{Behnke:2013xla,Baer:2013cma,Agashe:2013hma,Asner:2013psa,Moortgat-Picka:2015yla,Fujii:2015jha}.
  
%%%%%%%%%%%%%%%%%%%%%%%%%
\subsection{Correlations in Down - Charged lepton type models: $\boldsymbol{\thc}$ versus
$\boldsymbol{\thu}$\label{sSEC:thc_vs_thu}}
%%%%%%%%%%%%%%%%%%%%%%%%%
We have chosen down - charged lepton type models in order to have FCNC in the up sector while avoiding flavour violation in the charged lepton sector. 
In the down-type BGL model $d_\gamma$, with HFVNC in the up sector mediating rare top decays, following \refEQ{EQ:BRthq:01}, we get the simple correlation formulas
\begin{equation}
\frac{\text{Br}_{(d_{\gamma})}(\thc)}{\text{Br}_{(d_{\gamma})}(\thu)}=\frac{\abs{\V{c\gamma}}^{2}}{\abs{\V{u\gamma}}^{2}}\,.
\label{EQ:thq:ratio:01}
\end{equation}
Values of the ratio in \refEQ{EQ:thq:ratio:01} for different models are shown in table \ref{TAB:thq:ratio:CKMs}.
\begin{table}[h]
\begin{center}
\begin{tabular}{|c||c|c|c|}
\hline
Model $d_\gamma$& $d$ & $s$ & $b$ \\ \hline
$\frac{\abs{\V{c\gamma}}^2}{\abs{\V{u\gamma}}^2}$ & $\frac{\abs{\V{cd}}^2}{\abs{\V{ud}}^2}(\sim \lambda^{2})=0.0534$ & $\frac{\abs{\V{cs}}^2}{\abs{\V{us}}^2}(\sim \frac{1}{\lambda^{2}})=18.664$ & $\frac{\abs{\V{cb}}^2}{\abs{\V{ub}}^2}(\sim \frac{1}{\lambda^{2}(\rho^2+\eta^2)})=119.53$\\ \hline
\end{tabular}
\caption{Ratio ${\text{Br}_{(d_{\gamma})}(\thc)}/{\text{Br}_{(d_{\gamma})}(\thu)}$ for different down-type models; $\rho$ and $\eta$ are the Wolfenstein parameters \cite{Wolfenstein:1983yz}.\label{TAB:thq:ratio:CKMs}}
\end{center}
\end{table}

From Table \ref{TAB:thq:ratio:CKMs} one can read that, in models of types $s$ and $b$, it turns out that
 $\text{Br}_{(d_{\gamma})}(\thc)\geq \text{Br}_{(d_{\gamma})}(\thu)$ while in models $s$ we have
 the more exotic relation $\text{Br}_{(d_{\gamma})}(\thc)\leq \text{Br}_{(d_{\gamma})}(\thu)$.
\begin{figure}[h!]
\begin{center}
\includegraphics[width=0.5\textwidth]{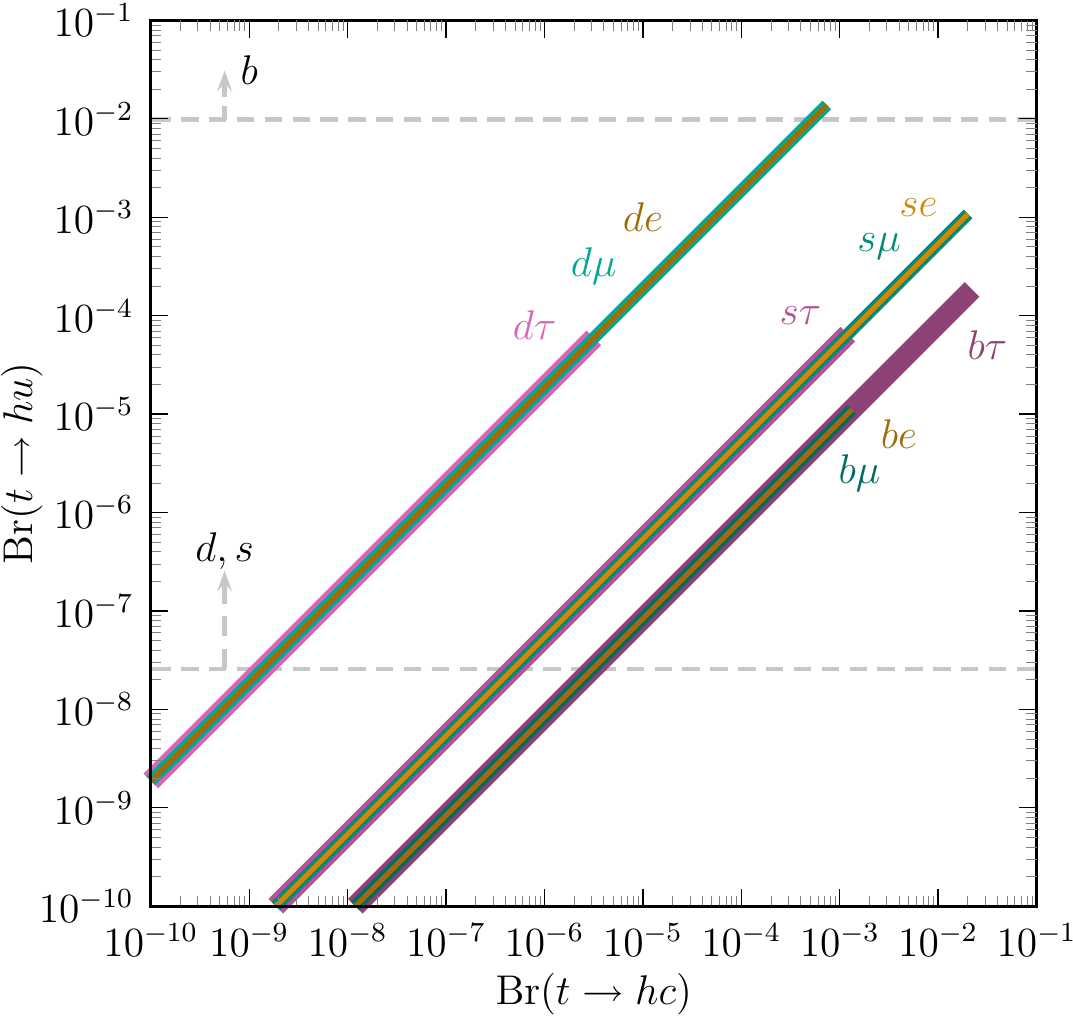}
\caption{Correlations in flavour changing decays $\thu$ and $\thc$ in models of type down quark - charged lepton, i.e. $(d_i,\ell_j)$. Line widths have no intrinsic meaning, they are only intended to help in distinguishing among the different models represented.\label{FIG:thuthc:downlep:01}}
\end{center}
\end{figure}
The correlated allowed ranges for the
branching ratios of these rare decay modes are shown in figure \ref{FIG:thuthc:downlep:01}.
The correlations -- the lines -- are associated with each particular model; for
example, the purple line labeled by the letters $b$ and $\tau$ 
is the correlation among the branching ratios $\text{Br}_{(b,\tau)}(\thu)$ and
 $\text{Br}_{(b,\tau)}(\thc)$, where the subscript $(b,\tau)$ identifies completely the model
 -- both the quark and the lepton type --. In that particular model,
 $\text{Br}_{(b,\tau)}(\thc)$ can reach values up to a few times $10^{-2}$,
 that is up to the actual experimental upper bounds.
 In models $(b,e)$ and $(b,\mu)$ -- in yellow and green respectively -- 
the correlations among $\text{Br}(\thu)$ and $\text{Br}(\thc)$ overlap along the same line with
 model $(b,\tau)$, because all three models share the same $\CKM$ factors. The only difference
 between $(b,\tau)$ and $(b,\mu)$ models in figure \ref{FIG:thuthc:downlep:01}, is in the maximum allowed value
of the factor $\ctti$, whose origin is in the different predictions for the processes involving flavour
conserving leptonic processes like $\nh\to \tau\bar\tau$ -- as considered in appendix \ref{AP:HiggsLHC} --.
 Without taking into account flavour changing low energy constraints, it is clear that the models $(b,\tau)$, 
$(d,e)$, $(d,\mu)$, $(s,e)$ and $(s,\mu)$, are the most promising models to discover new physics either in $\thc$ or $\thu$. 
These models have flavour changing couplings in the up sector, therefore the
Higgs coupling to $u\bar c$ could generate $D^{0}$ -- $\bar{D}^{0}$ mixing \cite{Blankenburg:2012ex}.
 To avoid too large $D^{0}$ -- $\bar D^{0}$ mixing induced by tree level Higgs exchange, one can naively obtain
the upper bounds for $\ctti$ collected in table \ref{TAB:D0mix:ctti}.

\begin{table}[h]
\begin{center}
\begin{tabular}{|c||c|c|c|}
\hline
Model $d_\gamma$& $d$ & $s$ & $b$ \\ \hline
$\ctti$ & $\leq 0.05$ & $\leq 0.05$ & $\leq 72.3$\\ \hline
\end{tabular}
\caption{Naive $D^0$ -- $\bar D^0$ constraint on $\ctti$ from tree level $\nh$ exchange.\label{TAB:D0mix:ctti}}
\end{center}
\end{table}

The potential effects of these constraints are represented in figure \ref{FIG:thuthc:downlep:01} with
exclusion horizontal dashed lines for the corresponding models.
 The potential constraint in $b$-type models is irrelevant\footnote{Only for $b\tau$ models, where the top quark decay constraints $|\ctti|<4.9$, \refEQ{eq:BRthq:ctti:01}, can be saturated, perturbative unitarity requirements may be relevant (see appendix \ref{AP:Pert}).},
 while this constraint could be more relevant in $s$ and $d$ models.
 Nevertheless, in these models we do not have just the 125 GeV Higgs boson,
 but in addition another scalar $\nH$ and a pseudoscalar $\nA$.
 It is well known that the scalar and pseudoscalar contributions to $D^{0}$ -- $\bar D^{0}$
 mixing cancel exactly in the limit of degenerate masses \cite{Nebot:2015wsa}. 
 Note that the contributions to the oblique parameters \cite{Peskin:1991sw} do also cancel in the
limit of degenerate masses and no mixing between the standard Higgs and the
other neutral scalars \cite{Grimus:2008nb}. These considerations imply that
one cannot translate into direct constraints the naive requirements on $\nh$ couplings,
since they can be relaxed in the presence of the other Higgses $\nH$ and $\nA$. Although
it is not within the scope of this paper to perform a complete analysis including
 the effects of the additional scalars, we illustrate how these
cancellations operate in the case of meson mixing constraints, in Appendix \ref{AP:Mix}, and for the constraints coming from $\mu
\to e\gamma$, where the cancellations are not so evident, in Appendix \ref{AP:MEG}. In the following,
 potential low energy flavour changing constraints are shown in the figures in the same fashion as in figure \ref{FIG:thuthc:downlep:01}.
It is important to keep in mind that they are indicative of which models are under pressure or else safer from that point of view.

The correlations in figure \ref{FIG:thuthc:downlep:01} correspond to the models of type down quark - charged
lepton, $(d_i,\ell_j)$, where FCNC are present in the up quark sector and in the neutrino sector. In these
models, FCNC are proportional to neutrino masses and thus there are no
flavour changing constraints coming from the leptonic sector. The
constraints from the Higgs signals involve the diagonal couplings which do change and were taken into
account as explained in appendix \ref{AP:HiggsLHC}. When down quark-neutrino type models are considered, $|\ctti|$ is also constrained by $\mu\to e\gamma$, $\tau\to e \gamma$, $\tau\to\mu\gamma$ and other flavour violating processes. It can be shown that, in any $\nu_i$ model, $\mu\to e\gamma$ is the most constraining process as far as $\ctti$ is concerned. We address in more detail the $\mu\to e\gamma$ restrictions in section \ref{sSEC:hmt_vs_hbqthq} and in appendix \ref{AP:MEG}.
% In the case of models of type down quark - neutrino, there are also constraints on $\ctti$ coming from $\mu \to e\gamma $, which are addressed in more detail in section \ref{sSEC:hmt_vs_hbqthq} and in appendix \ref{AP:MEG}.

%%%%%%%%%%%%%%%%%%%%%%%%%
\subsection{Correlations in Up-Charged lepton models: $\boldsymbol{\hbs}$ versus $\boldsymbol{\hbd}$\label{sSEC:hbs_vs_hbd}}
%%%%%%%%%%%%%%%%%%%%%%%%%

\begin{figure}[h!]
\begin{center}
\includegraphics[width=0.5\textwidth]{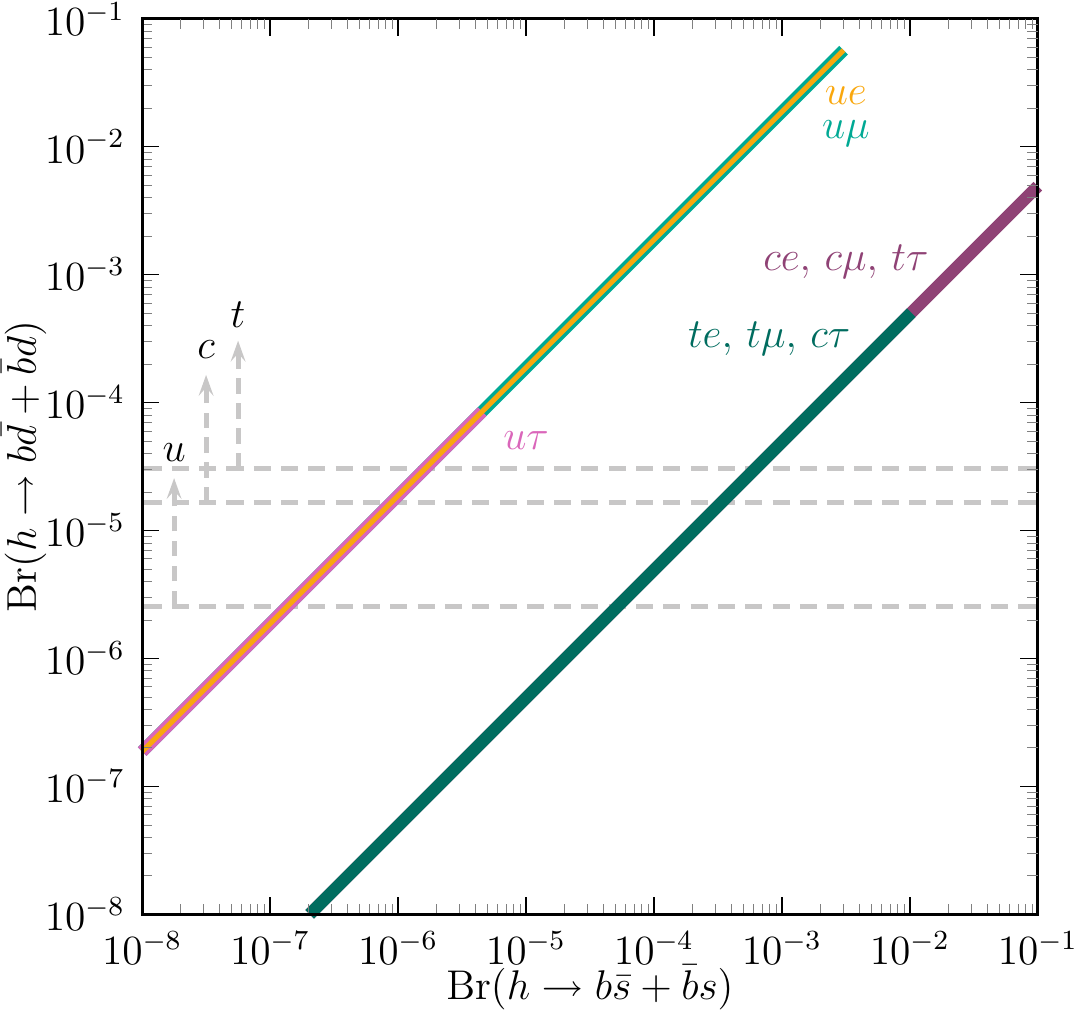}
\caption{Correlations in flavour changing $\nh$ decays to down quarks in $(u_i,\ell_j)$ models.\label{FIG:hbdhbs:uplep:01}}
\end{center}
\end{figure}

In order to have $\hbq$ decays at a significant rate,
we have to consider up-type models, $u_{k}$,
where FCNC occur in the down sector. In this case, following \refEQS{EQ:Gamma:hbq:01}-\eqref{EQ:Gamma:hbq:02}, the correlations among the Higgs flavour changing decays to down quarks are given by
\begin{equation}
\frac{\text{Br}_{(u_k)}( \nh\to b\bar s+s\bar b)}{\text{Br}_{(u_k)}( \nh\to b\bar d+d\bar b)}=
\frac{\abs{\V{ks}}^{2}}{\abs{\V{kd}}^{2}}\,.
\label{EQ:hbs:hbd:01}
\end{equation}
The values of the ratio in \refEQ{EQ:hbs:hbd:01} for the different up-type models are given in table \ref{TAB:hbq:CKM:02};
 the correlations are represented in figure \ref{FIG:hbdhbs:uplep:01}.
\begin{table}
\begin{center}
\begin{tabular}{|c||c|c|c|}
\hline
Model $u_k$ & $u$ & $c$ & $t$ \\ \hline
$\frac{\abs{\V{ks}}^2}{\abs{\V{kd}}^2}$ & $\frac{\abs{\V{us}}^2}{\abs{\V{ud}}^2}(\sim \lambda^{2})=0.0535$ & $\frac{\abs{\V{cs}}^2}{\abs{\V{cd}}^2}(\sim \frac{1}{\lambda^{2}})=18.688$  & $\frac{\abs{\V{ts}}^2}{\abs{\V{td}}^2}(\sim \frac{1}{\lambda^{2}})=20.409$\\ \hline
\end{tabular}
\caption{$\CKM$ factors entering \refEQ{EQ:hbs:hbd:01}.\label{TAB:hbq:CKM:02}}
\end{center}
\end{table}

The correlations in figure \ref{FIG:hbdhbs:uplep:01} follow from the full data
analysis in appendix \ref{AP:HiggsLHC}, including the necessary study of the $\nh$ total decay width in these BGL models.
 We can see that in $(t,\tau)$, $(c,e)$ and $(c,\mu)$ models, the channel $\nh\to b\bar s+\bar bs$ can have a branching ratio at the $10^{-1}$ level,
 while in $(c,\tau)$, $(t,e)$ and $(t,\mu)$ models this branching ratio can be near $10^{-2}$ values. It is also 
remarkable that in $(u,e)$ and $(u,\mu)$ models, the branching ratio of $\nh\to b\bar d+\bar bd$
 can also reach values not far from the $10^{-1}$ level. Top endpoints of the correlations follow from the general analysis,
 which allows maximal values $\abs{\ctti}\sim 8$. 
As in the case of $\thq$ correlations, the dashed lines show
 the naive constraints one would get from $\nh$ contributions alone to the low energy processes $K^0$ -- $\bar K^0$,
 $D^0$ -- $\bar D^0$, $B_d^0$ -- $\bar B_d^0$ and $B_s^0$ -- $\bar B_s^0$; explicit values are collected in table \ref{TAB:P0mix:ctti}.
 Examples of models where low energy contraints can be relevant are models $(u,e)$ and $(u,\mu)$. Once again, we must stress that the presence of other contributions in these BGL models can relax these low energy constraints. We include them here for the sake of completeness.

\begin{table}[h]
\begin{center}
\begin{tabular}{|c||c|c|c|}
\hline
Model $u_k$& $u$ & $c$ & $t$ \\ \hline
$\ctti$ & $\leq 0.43$ & $\leq 0.43$ & $\leq 0.60$\\ \hline
\end{tabular}
\caption{Naive $K^0$ -- $\bar K^0$, $B_d^0$ -- $\bar B_d^0$ or $B_s^0$ -- $\bar B_s^0$ constraint on $\ctti$ from tree level $\nh$ exchange.\label{TAB:P0mix:ctti}}
\end{center}
\end{table}
 Note, however, that since the values in Table \ref{TAB:P0mix:ctti} are near $1$, one cannot go too far above the dashed lines in figure \ref{FIG:hbdhbs:uplep:01} without taking into account perturbative unitarity (see appendix \ref{AP:Pert}).

%%%%%%%%%%%%%%%%%%%%%%%%%
\subsection{Correlations in neutrino models: $\boldsymbol{\hmt}$ together with $\boldsymbol{\thq}$ or $\boldsymbol{\hbq^\prime}$\label{sSEC:hmt_vs_hbqthq}}
%%%%%%%%%%%%%%%%%%%%%%%%%

In neutrino models, we have flavour changing Higgs interactions in the leptonic
sector giving rise to the interesting processes $\nh\to \mu^\pm \tau^\mp$, $e^\pm\tau^\mp$, $e^\pm\mu^\mp$.
 The corresponding couplings are proportional to one of the lepton masses, therefore the transitions
including a $\tau$ lepton are at least more probable by a factor $(m_{\tau}/m_{\mu})^{2}$.
 We will concentrate in these transitions, containing a $\tau$ lepton, even if experimentally the $\mu e$
channel is very interesting. These transitions are also proportional to $\CTTI$, like all
Higgs induced flavour changing transitions in these models; therefore, in down-type
models, we will have perfectly defined correlations between $\hmt$ and $\thq$; in up-type models, we will have
correlations among $\hmt$ and $\hbq$.

At present, as already mentioned, evidence from the CMS collaboration \cite{Khachatryan:2015kon}
points towards the possible observation of the decay $\hmt$, which would definitely be ``beyond the SM physics''.
 These predictions could be checked by looking at the correlations with the channels  $\thq$ for
down-type BGL models, and with $\hbq$ in up-type BGL models.

\begin{figure}[htb]
\begin{center}
\includegraphics[width=0.5\textwidth]{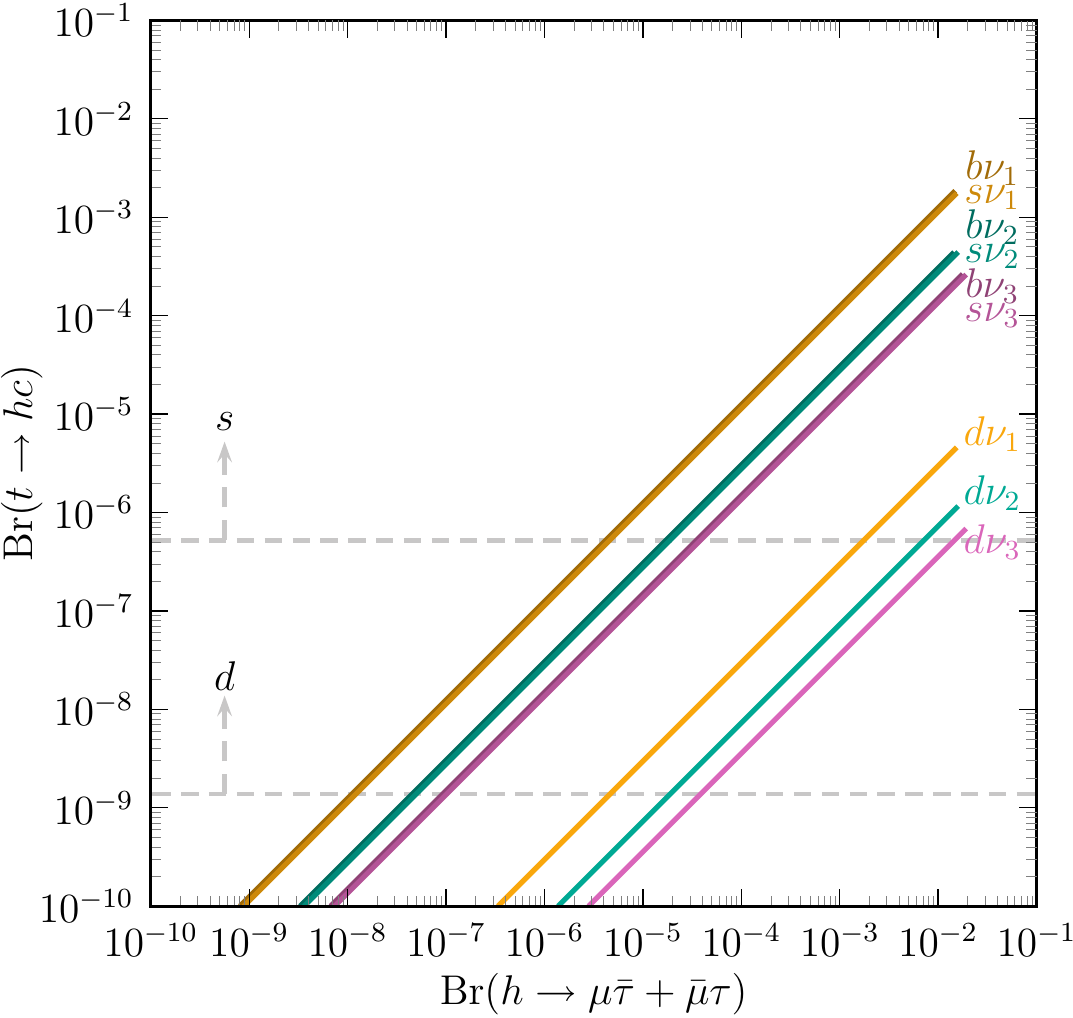}
\caption{Correlations in flavour changing $\thc$ vs. $\hmt$ decays in $(d_\gamma,\nu_\sigma)$ models.\label{FIG:thchmt:downnu:01}}
\end{center}
\end{figure}

In BGL models of $(d_\gamma,\nu_\sigma)$ type, the correlations $\thq$
versus $\nh\to \mu \bar{\tau}+\tau\bar{\mu}$, following \refEQS{EQ:BRthq:01}-\eqref{EQ:hmt:Gamma:01}, are
controlled by 
\begin{equation}
\text{Br}_{(d_{\gamma})}(\thq)
=2.063\,\left|\frac{\V{q\gamma}\V{t\gamma}}{\V{tb}\,\U{\mu\sigma}\U{\tau\sigma}}\right|^{2}
\,\frac{\Gamma_\nh}{\Gamma^{\rm[SM]}_\nh}\,\text{Br}_{(\nu_{\sigma})}(\nh\to \mu \bar{\tau}+\tau\bar{\mu}) 
\end{equation}
Notice that these correlations are fixed by CKM and PMNS matrix elements.
Nevertheless, there is also the ratio of the total width of the Higgs in BGL models
versus the SM value. This ratio makes the correlation to depart from strict linearity depending
on $\cab$ and $\tb$. In figure \ref{FIG:thchmt:downnu:01}, we first show the plot
where only the range of variation is displayed, that is the strict linear relation.
In this plot, figure \ref{FIG:thchmt:downnu:01}, one can see the effect of the upper bound on
 $\text{Br}_{(\nu_{\sigma})}(\nh\to\mu\bar{\tau}+\tau\bar{\mu})$. In particular,
 in models $(b,\nu_{1})$, the maximum value that $\text{Br}_{(b,\nu_{1})}(\thc)$ may reach is a few times $10^{-3}$.
 This value is smaller than the maximum allowed value in $(b,\tau)$ models, presented in figure \ref{FIG:thuthc:downlep:01}.
 As usual, we have included -- with dashed lines -- the naive constraints coming from the individual Higgs contribution to low energy flavour
changing hadronic processes.

\begin{figure}[htb]
\begin{center}
\includegraphics[width=0.5\textwidth]{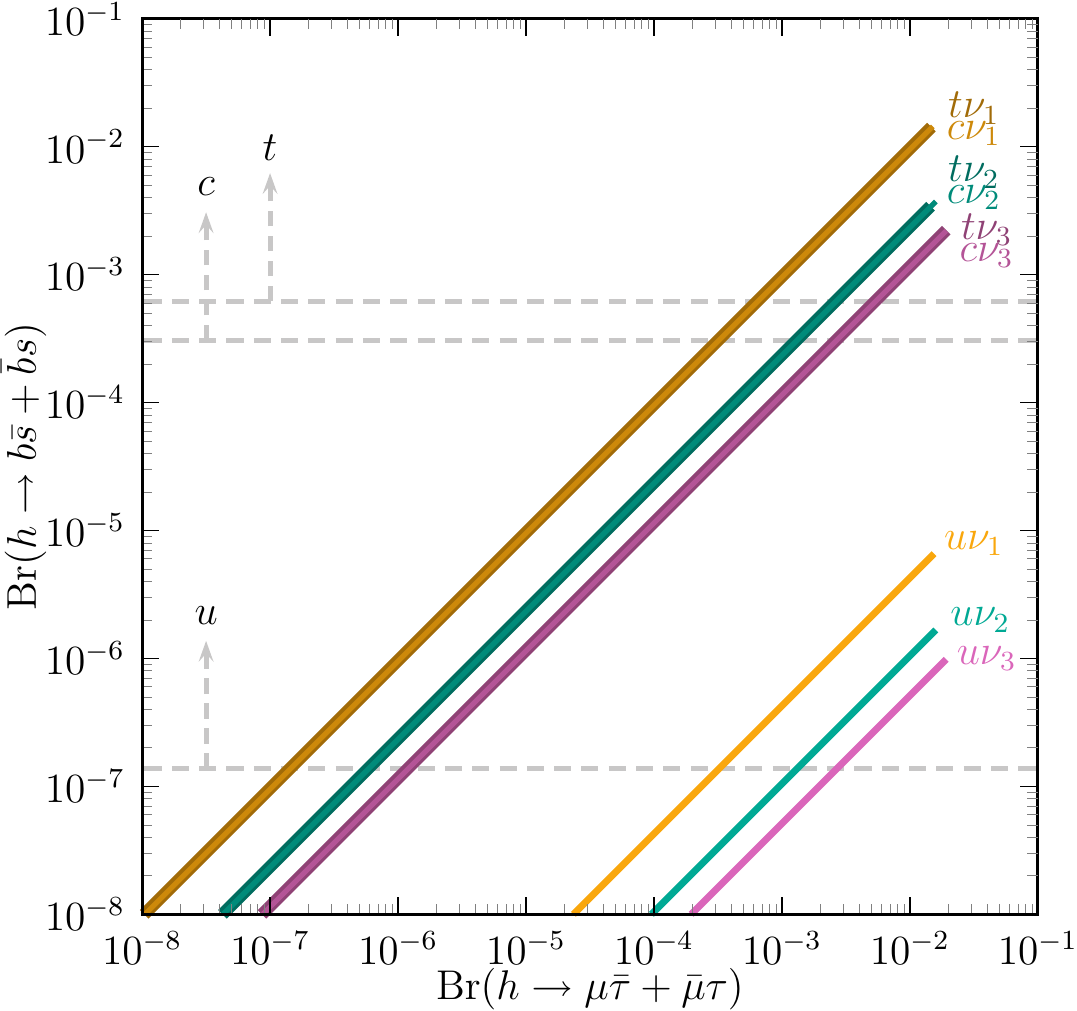}
\caption{Correlations in flavour changing $\hbs$ vs. $\hmt$ decays in $(u_\gamma,\nu_\sigma)$ models.\label{FIG:hbshmt:upnu:01}}
\end{center}
\end{figure}

In BGL models of $(u_k, \nu_\sigma)$ type, we have similar expressions for the
correlation among $\hbq$ and $\nh\to\mu\bar{\tau}+\tau\bar{\mu}$ decays.
 The corresponding plot is shown in figure \ref{FIG:hbshmt:upnu:01}.
We can observe again the effects of the measurement in the $\hmt$ channel. Nevertheless, as one can see,
$\nh\to b\bar{s}+s\bar{b}$ branching ratios can still have values above the $10^{-2}$ level.

\begin{figure}[htb]
\begin{center}
\subfigure[Without $\mu\to e\gamma$ bound.\label{FIG:thchmt:s1s3:01a}]{\includegraphics[width=0.4\textwidth]{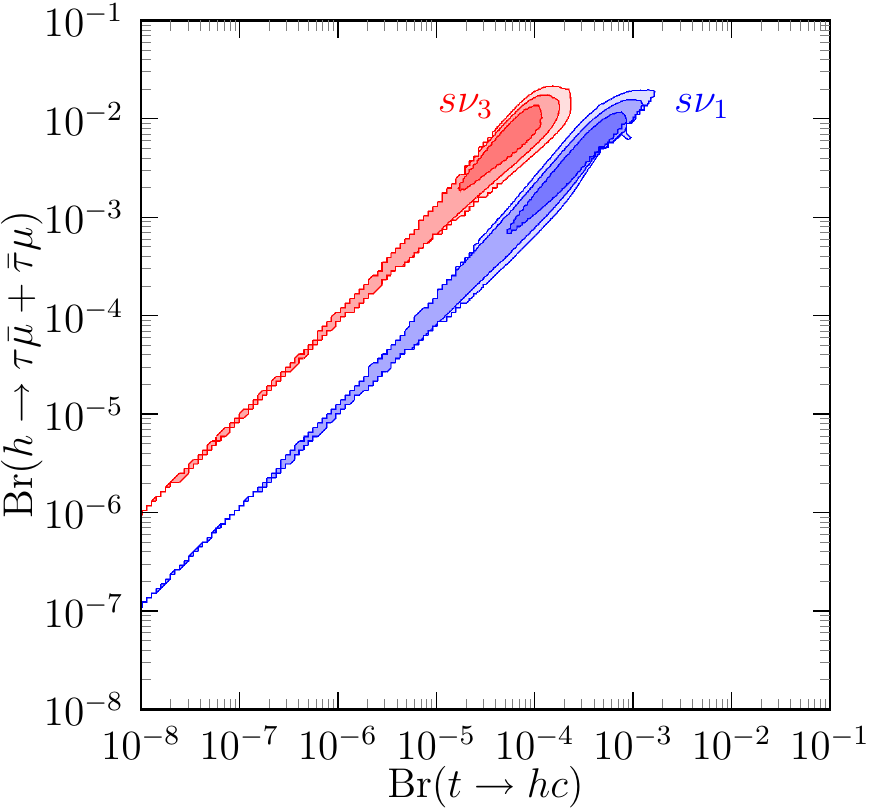}}\qquad
\subfigure[With $\mu\to e\gamma$ bound.\label{FIG:thchmt:s1s3:01b}]{\includegraphics[width=0.4\textwidth]{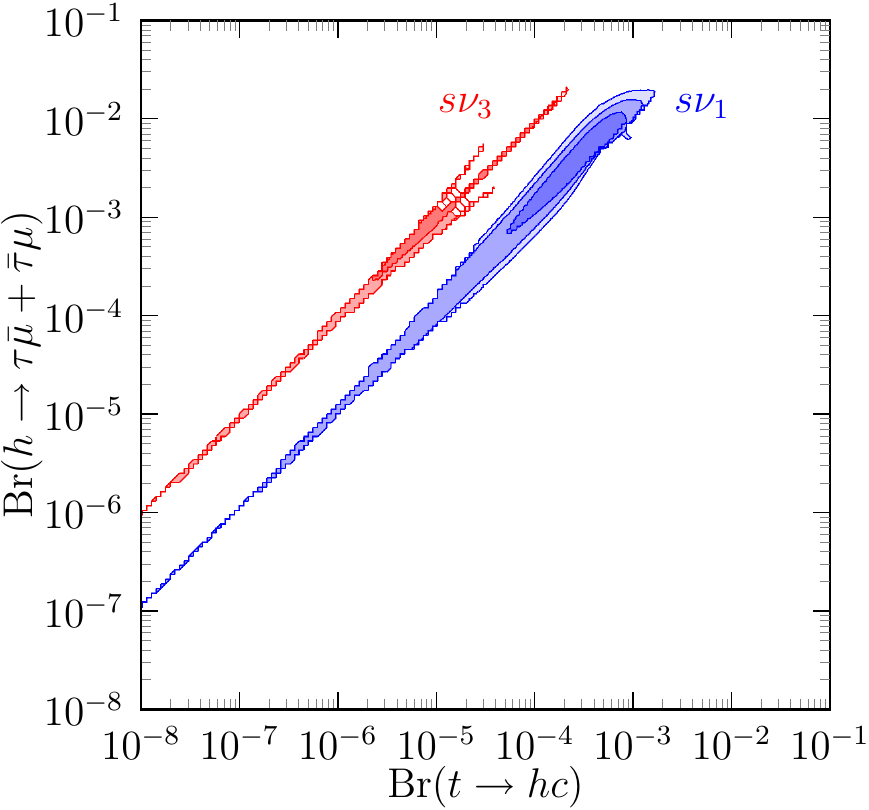}}
\caption{Correlations in flavour changing $\thc$ vs. $\hmt$ decays in $(s,\nu_1)$ and $(s,\nu_3)$ models. The regions, darker to lighter, correspond to $1$, $2$ and $3\sigma$ regions from the full analysis. As anticipated, they are not simple straight lines. Furthermore, blue and
red regions here correpond to yellow and purple lines in
figure \ref{FIG:thchmt:downnu:01}. The $1\sigma$ region reflects the effect of the CMS measurement, which is compatible with zero at the $2.4\sigma$ level.\label{FIG:thchmt:s1s3:01}}
\end{center}
\end{figure}

\begin{figure}[htb]
\begin{center}
\includegraphics[width=0.4\textwidth]{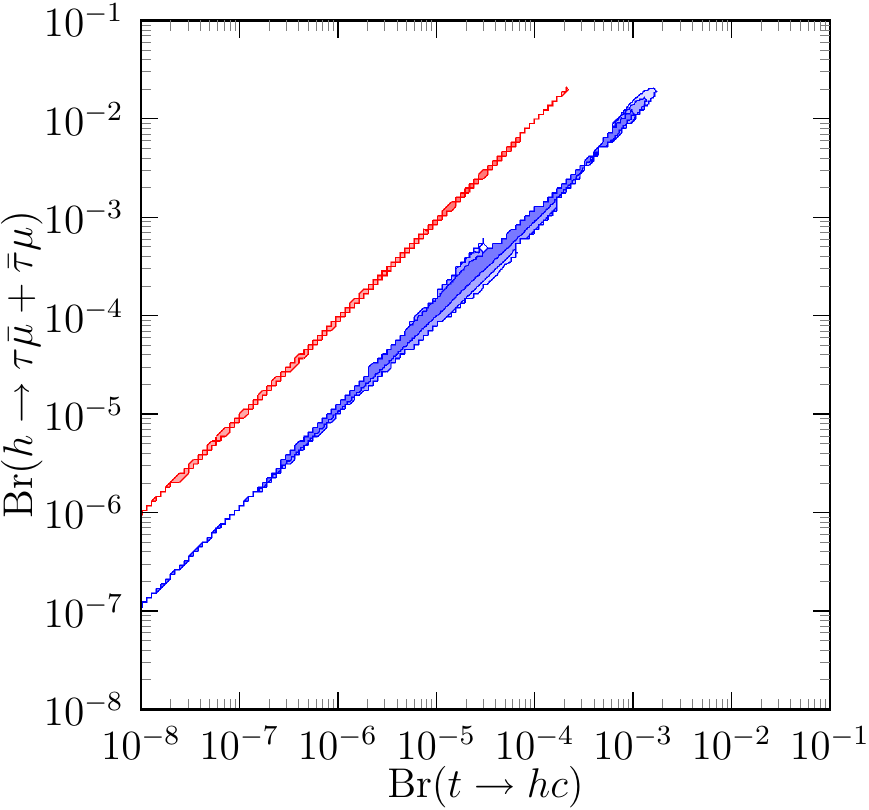}
\caption{Correlations in flavour changing $\thc$ vs. $\hmt$ decays in $(s,\nu_1)$ and $(s,\nu_3)$ models with $\mu\to e\gamma$ following scenario (2) (see text and caption of figure \ref{FIG:thchmt:s1s3:01}).\label{FIG:thchmt:s1s3:02}}
\end{center}
\end{figure}

Although Figures \ref{FIG:thchmt:downnu:01} and \ref{FIG:hbshmt:upnu:01} show $\nh\to\mu\tau$ decays, the values corresponding to $\nh\to e\tau$ decays follow from the PMNS factors in Table \ref{TAB:hmt:PMNS}. Notice that for $\nh\to e\mu$ decays, an additional suppression factor $(m_\mu/m_\tau)^2\simeq 3.5\times 10^{-3}$ is involved. 
It is important to stress that perturbative unitarity will not impose any further constraint on Figures \ref{FIG:thchmt:downnu:01} and \ref{FIG:hbshmt:upnu:01} because \refEQ{eq:ctti_hmt:00} is at work. 
Several authors  have noticed that $\mu\to e\gamma $ constrains very
severely the coupling $\nh\to \mu e$ via the mass unsuppressed two-loop Barr-Zee diagrams \cite{Barr:1990vd}.
 Since in BGL models all leptonic flavour changing
Higgs couplings are fixed by $\PMNS$, masses and $\ctti$, it is clear that the $\mu\to e\gamma$ bound will
translate into an important constraint on $\ctti$, that has to be incorporated to the global
analysis. However, in these two-loop diagrams -- as in the case of the different neutral
meson mixings --, not only the Higgs $\nh$ can be exchanged, but also the other scalar $\nH$
and pseudoscalar $\nA$ will enter with the possibility to produce destructive interference between the
different contributions. As detailed in appendix \ref{AP:MEG}, we have made a global analysis fully incorporating the different
contributions to $\mu \to e\gamma $ (in BGL models) in two different cases: (1) with
varying free values of the masses $m_{\nH}$ and $m_{\nA}$ below 1 TeV, and (2) by imposing 
$m_{\nH}-m_{\nA}\leq 50$ GeV and varying $m_{\nH}$ below 1 TeV: in both scenarios,
oblique corrections can be maintained under control.
To illustrate how these cancellations operate in $\mu \to e\gamma $, we
represent the correlation among $\nh\to\mu\bar{\tau}+\tau\bar{\mu}$ and $\thc$ in models $(s,\nu_{3})$ and $(s,\nu_{1})$.
 In figure \ref{FIG:thchmt:s1s3:01} we show this correlation in the full analysis, first without including the $\mu\to e\gamma $ constraint, figure \ref{FIG:thchmt:s1s3:01a},
and then, in figure \ref{FIG:thchmt:s1s3:01b}, when we introduce the $\mu \to e\gamma $ constraint as mentioned in
scenario (1), that is with free values of $m_{\nH}$ and $m_{\nA}$ below 1 TeV.
As figure \ref{FIG:thchmt:s1s3:01} shows, the region of variation of the correlation remains
essentially the same, meaning that there are cancellations at work,
implying that in this kind of 2HDM, one cannot forget about the additional Higgses
in order to impose the low energy constraints.
Considering instead scenario (2), i.e. taking $m_{\nH}-m_{\nA}\leq 50$ GeV and varying $m_{\nH}$ below $1$ TeV, 
 the corresponding plot is shown in figure \ref{FIG:thchmt:s1s3:02}.

It is then clear that if we include in the analysis the complete two-loop Bar-Zee contribution to
 $\mu \to e\gamma$, we can conclude from the observed changes and the
actual level of precision, that effects are not yet relevant in the
majority of BGL models. We have illustrated this result with down-type models,
but the same happens in up-type models; therefore, $\hbs$
correlations with $\hmt$ remain essentially unchanged without the inclusion of the $\mu \to e\gamma $ constraint.

\clearpage
 
%%%%%%%%%%%%%%%%%%%%%%%%%%%%%%%%%%%%%%%%
\section{Conclusions\label{SEC:Conclusions}}
%%%%%%%%%%%%%%%%%%%%%%%%%%%%%%%%%%%%%%%%
We analyse flavour changing scalar couplings in the framework of a class of two Higgs Doublet models where these couplings arise at tree level, but with their flavour structure entirely determined by the CKM and PMNS matrices. This very special structure of the scalar couplings is stable under the renormalization group, since it results from a discrete symmetry of the Lagrangian. The symmetry can be implemented in various ways, corresponding to a variety of BGL models. We pointed out that this class of models leads to New Physics with potential for being discovered at the LHC and/or at an ILC. We examine in detail rare top decays like $\thq$ $(q=u,c)$ and leptonic flavour changing decays such as $\nh\to\tau^\pm\ell^\mp$ $(\ell=\mu,e)$, as well as hadronic flavour changing decays like $\hbs$ and $\hbd$. All these decays occur in the SM only at loop level and therefore are strongly suppressed. In BGL models, the flavour violating couplings occur at tree level, but some of the most dangerous couplings are suppressed by small CKM elements. We address the question whether there are regions in some of the BGL models where these couplings are such that may lead to the discovery of rare flavour violating processes at the LHC-13TeV.
 
We also do a systematic study of the correlations among various observables which are an interesting distinctive feature of BGL models. In the search of these regions, we have taken into account the low energy restrictions on flavour violating processes as well as the stringent constraints on all SM processes associated to the Higgs production and subsequent decays in the various channels.

As far as the low energy flavour constraints are concerned, we agree with other authors that these cannot be imposed by assuming the dominance of the lighter Higgs contribution. This was known, in particular, in BGL models for neutral Meson-Antimeson mixing: there are important cancellations among different virtual Higgs contributions. We have illustrated this point showing how these cancellations operate in the two-loop, Higgs mediated, $\mu\to e\gamma$ process, where these cancellations can appear in the amplitude, operating at the level of one or two orders of magnitude.

Two Higgs doublet models are one of the simplest extensions of the the SM. In general, they have a large number of free parameters and lead to scalar FCNC which have severe restrictions from low energy flavour violating processes.
BGL models have the notable feature of having a small number of free parameters and achieving a natural suppression of these couplings, while at the same time allowing for the exciting scenario of having some flavour violating top and Higgs decays to occur at discovery level at the LHC-13TeV. 

\section*{Acknowledgments}
The authors thank Andre David for useful discussions on LHC Higgs 
 results, Joan Fuster and Christophe Grojean for interesting 
 discussions, and Marc Sher for helpful comments. This work is partially supported by Spanish MINECO under grants FPA2011-23596, FPA2015-68318-R and SEV-2014-0398,
by Generalitat Valenciana under grant GVPROMETEOII 2014-049
and by Funda\c{c}\~ao para a Ci\^encia e a
Tecnologia (FCT, Portugal) through the projects CERN/FP/123580/2011,
PTDC/FIS-NUC/0548/2012, and CFTP-FCT Unit 777 
(UID/FIS/00777/2013) which are partially funded through POCTI (FEDER),
COMPETE, QREN and EU. MN is supported by a postdoctoral fellowship from
Project PTDC/FIS-NUC/0548/2012. Part of this work was done at the Theory Division 
of CERN, where GCB spent time as Scientific Associate and the remaining authors as Short 
Term Visitors. The authors also acknowledge the hospitality of Universidad de Valencia, IFIC,
and CFTP at IST Lisboa during visits for scientific collaboration. 

\clearpage
\appendix

%%%%%%%%%%%%%%%%%%%%%%%%%%%%%%%%%%%%%%%%
\section{Higgs signals\label{AP:HiggsLHC}}
%%%%%%%%%%%%%%%%%%%%%%%%%%%%%%%%%%%%%%%%
Besides the appearance of flavour changing couplings of the Higgs boson, as shown in equations \REFEQS{EQ:QuarkYukawas:01}--\eqref{yuij}, flavour conserving couplings are modified owing to the mixing in the scalar sector, and thus a detailed analysis of the constraints on $\alpha-\beta$ and $\tan\beta$ that Higgs data impose is mandatory. The experimental information concerning the SM-like with mass 125 GeV discovered at the LHC is summarised in a set of signal strengths of the form
\begin{equation}
\sigstr{i}{X}=\frac{\left[\sigma(pp\to \nh)\right]_i}{\left[\sigma(pp\to \nh)_{\rm SM}\right]_i}\frac{\text{Br}(\nh\to X)}{\text{Br}(\nh\to X)_{\rm SM}}
\end{equation}
where $i$ labels the different combinations of production mechanisms and $X$ the decay channels. Concerning production, the most relevant modes \cite{Heinemeyer:2013tqa} are gluon-gluon fusion (ggF), vector boson fusion (VBF) and Higgstrahlung (WH \& ZH); values used in the analysis are collected in Table \ref{TAB:production}.
\begin{table}[h!]
\begin{center}
\begin{tabular}{|c||c|c|c|c|c|c|}
\hline
Production channel $i$ & ggF & VBF & $W\nh$ & $Z\nh$ & $t\bar t\nh$ & $b\bar b\nh$ \\ \hline
$\left[\sigma(pp\to \nh)_{\rm SM}\right]_i$ (pb) & 19.27 & 1.578 & 0.7046 & 0.4153 & 0.1293 & 0.2035 \\ \hline
\end{tabular}
\caption{Production cross sections (uncertainties are not shown), for $m_{\nh}=125.0$ GeV and $\sqrt{s}=8$ TeV.}
\label{TAB:production}
\end{center}
\end{table}
Relevant branching ratios within the SM are in turn collected in Table \ref{TAB:decayBR}.
\begin{table}[h!]
\begin{center}
\begin{tabular}{|c||c|c|c|c|c|c|}
\hline
Decay channel $X$ & $b\bar b$ & $WW^\ast$ &  $ZZ^\ast$ & $\tau\bar\tau$ & $\gamma\gamma$ & $gg$ \\ \hline
$[\text{Br}(\nh\to X)]_{\rm SM}$ & 0.578 & 0.216 & 0.0267 & 0.0637 & 0.0023 & 0.0856 \\ \hline
\end{tabular}
\caption{Decay branching ratios for $m_{\nh}=125.0$ GeV (the total width is $\Gamma_\nh=4.03$ MeV).}
\label{TAB:decayBR}
\end{center}
\end{table}

We now list the different signal strengths obtained by the CMS and ATLAS collaborations organising them by decay channel.

\begin{itemize}
\item $\nh\to\gamma\gamma$,
\begin{equation}
\sigstr{ggF}{\gamma\gamma}=1.12\begin{smallmatrix}+0.37\\ -0.32\end{smallmatrix}\,,\qquad 
\sigstr{VBF}{\gamma\gamma}=1.58\begin{smallmatrix}+0.77\\ -0.68\end{smallmatrix}\,,\qquad 
\sigstr{VH}{\gamma\gamma}=-0.16\begin{smallmatrix}+1.16\\ -0.79\end{smallmatrix}\,,\qquad \text{CMS \cite{Khachatryan:2014ira}}
\end{equation}
\begin{equation}
\sigstr{ggF}{\gamma\gamma}=1.32\begin{smallmatrix}+0.41\\ -0.33\end{smallmatrix}\,,\qquad 
\sigstr{VBF}{\gamma\gamma}=0.78\begin{smallmatrix}+0.80\\ -0.64\end{smallmatrix}\,,\qquad \text{ATLAS \cite{Aad:2014eha}}.
\end{equation}

\item $\nh\to ZZ$ \cite{Chatrchyan:2013mxa},
\begin{equation}
\sigstr{0/1\,\text{jet}}{ZZ}=0.858\begin{smallmatrix}+0.321\\ -0.258\end{smallmatrix}\,,\qquad 
\sigstr{2\,\text{jet}}{ZZ}=1.235\begin{smallmatrix}+0.852\\ -0.583\end{smallmatrix}\,,
\end{equation}
where the dominant production in the 0/1 jet signal is gluon-gluon fusion -- ggF:VBF $\sim$ 20:1 --, while in the 2 jets signal, although gluon-gluon fusion contributes the most, VBF and VH are not negligible, -- ggF:VBF:VH $\sim$ 4:2:1 --.

\item $\nh\to WW$ \cite{Chatrchyan:2013iaa},
\begin{equation}
\sigstr{0/1\,\text{jet}}{WW}=0.74\begin{smallmatrix}+0.22\\ -0.20\end{smallmatrix}\,,\qquad 
\sigstr{\text{VBF+VH}}{WW}=0.60\begin{smallmatrix}+0.54\\ -0.46\end{smallmatrix}\,,
\end{equation}
where, as in $\mu^{ZZ}$, the 0/1 jet signal is gluon-gluon fusion dominated.

\item $\nh\to\tau\bar\tau$ \cite{Chatrchyan:2014nva},
\begin{equation}
\sigstr{0\,\text{jet}}{\tau\tau}=0.34\pm 1.09\,,\qquad \sigstr{1\,\text{jet}}{\tau\tau}=1.07\pm 0.46\,,\qquad \sigstr{2\,\text{jet-VBF}}{\tau\tau}=0.94\pm 0.41\,,
\end{equation}
where the 0 and 1 jet signals are dominated by gluon-gluon fusion and for the 2 jet-VBF tag, ggF and VBF production are similar.
\item $\nh\to\bar bb$ \cite{Chatrchyan:2013zna,Aad:2014xzb},
\begin{equation}
\sigstr{\text{VH}}{bb}=1.0\pm 0.5\,,\qquad \sigstr{\text{VH}}{bb}=0.65\begin{smallmatrix}+0.43\\ -0.40\end{smallmatrix}\,,
\end{equation}
\end{itemize}

Concerning the dependence of the couplings involved in the different production and decay channels, $\nh WW$ and $\nh ZZ$ are rescaled by a factor $\sab$ with respect to the SM for all the models. This affects VBF and VH production modes, $\nh\to WW,ZZ$ decays and the $W$-loop contribution to $\nh\to\gamma\gamma$, that we address later. For the couplings of $\nh$ to fermions, the picture is more involved, they are modified in a model dependent manner. We remind in Tables \ref{TAB:Htt}, \ref{TAB:Hbb} and \ref{TAB:HTT} the changes in $\nh\bar tt$, $\nh\bar bb$ and $\nh\bar\tau\tau$ with respect to the SM, where the $\nh \bar ff$ interaction is simply $\frac{m_f}{v}\nh\bar ff$.

\begin{table}[h!]
\begin{center}
\begin{tabular}{|c||c|c|c|}
\hline
Model type & $u,c$ & $t$ & Down model $d_i$ \\\hline
Coupling $m_t/v$ $\times$ & $\sab+\cab\tb$ & $\sab-\cab\tbinv$ & $\sab+\cab[(1-|\V{td_i}|^2)\tb-|\V{td_i}|^2\tbinv]$\\\hline
\end{tabular}
\caption{$\nh \bar tt$ coupling.\label{TAB:Htt}}\vspace{0.6cm}
\begin{tabular}{|c||c|c|c|}
\hline
Model type & $d,s$ & $b$ & Up model $u_i$ \\\hline
Coupling $m_b/v$ $\times$ & $\sab+\cab\tb$ & $\sab-\cab\tbinv$ & $\sab+\cab[(1-|\V{u_ib}|^2)\tb-|\V{u_ib}|^2\tbinv]$\\\hline
\end{tabular}
\caption{$\nh\bar bb$ coupling.\label{TAB:Hbb}}
\vspace{0.5cm}
\begin{tabular}{|c||c|c|c|}
\hline
Model type & $e,\mu$ & $\tau$ & Neutrino model $\nu_i$ \\\hline
Coupling $m_\tau/v$ $\times$ & $\sab+\cab\tb$ & $\sab-\cab\tbinv$ & $\sab+\cab[(1-|\V{\tau i}|^2)\tb-|\V{\tau i}|^2\tbinv]$\\\hline
\end{tabular}
\caption{$\nh\bar\tau\tau$ coupling\label{TAB:HTT}}
\end{center}
\end{table}

Notice that
\begin{itemize}
\item the change in $\nh\bar\tau\tau$ only affects the branching ratio in $\sigstr{i}{\tau\tau}$,
\item the change in $\nh\bar bb$ would in principle only affect the branching ratio in $\sigstr{i}{bb}$; however, production through the otherwise negligible $b\bar b\to\nh$ process could be $\tan\beta$ or $\tan^{-1}\beta$ enhanced.
\item the change in $\nh\bar tt$ affects gluon-gluon production and the top loop in $\nh\to\gamma\gamma$ decays.
\end{itemize}

The $\nh\to\gamma\gamma$ decay deserves some attention. In the SM it is a loop induced process where virtual $W$ and top diagrams interfere destructively. Besides the individual rescaling of both contributions, additional contributions mediated by the charged scalar $\cH$ could also contribute. Scenarios with sizable $\cH$ contributions to $\nh\to\gamma\gamma$ require a specific analysis that is beyond the scope of this work. A regime with heavy $\cH$ bosons can always be considered where this approximation is sound.

\begin{figure}[h!]
\begin{center}
% All in one row
% \subfigure[Model $t$ $\nu_2$]{\includegraphics[width=0.2\textwidth]{log10tb_ab_t2.png}}\qquad
% \subfigure[Model $d$ $\nu_1$]{\includegraphics[width=0.2\textwidth]{log10tb_ab_d2.png}}\qquad
% \subfigure[Model $u$ $e$]{\includegraphics[width=0.2\textwidth]{log10tb_ab_ue.png}}\qquad
% \subfigure[Model $b$ $\tau$]{\includegraphics[width=0.2\textwidth]{log10tb_ab_bt.png}}
% Or in 2 rows but larger
\subfigure[Model $t$ $\nu_2$]{\includegraphics[width=0.3\textwidth]{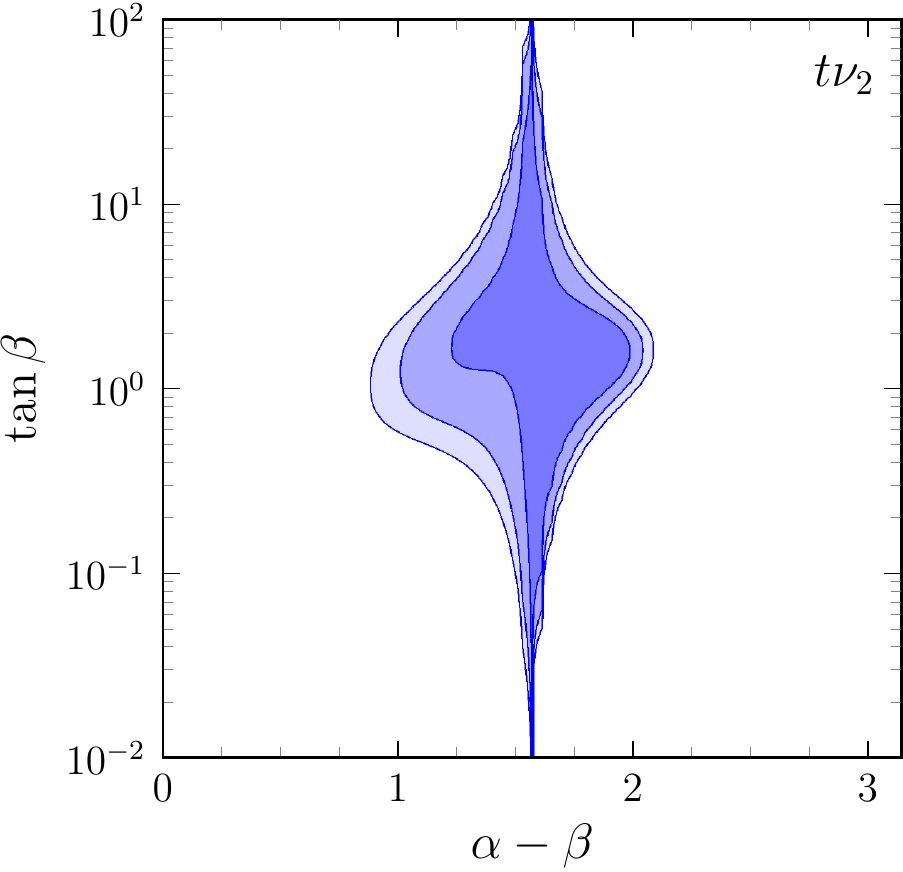}}\qquad
\subfigure[Model $d$ $\nu_1$]{\includegraphics[width=0.3\textwidth]{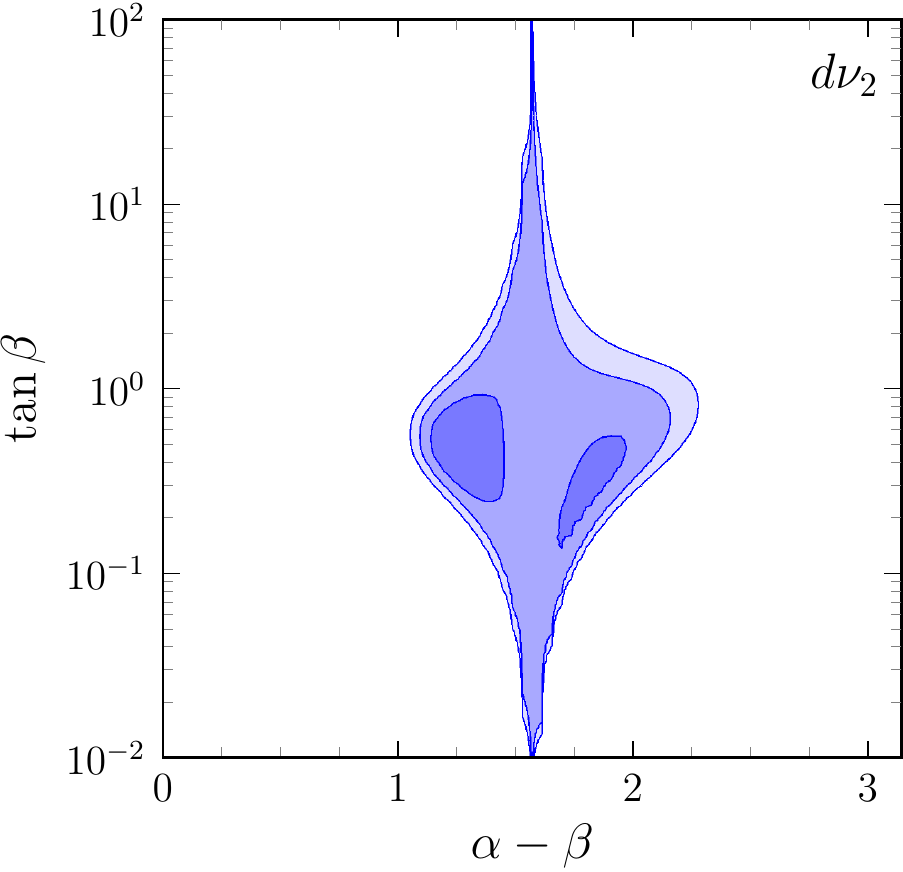}}\\
\subfigure[Model $u$ $e$]{\includegraphics[width=0.3\textwidth]{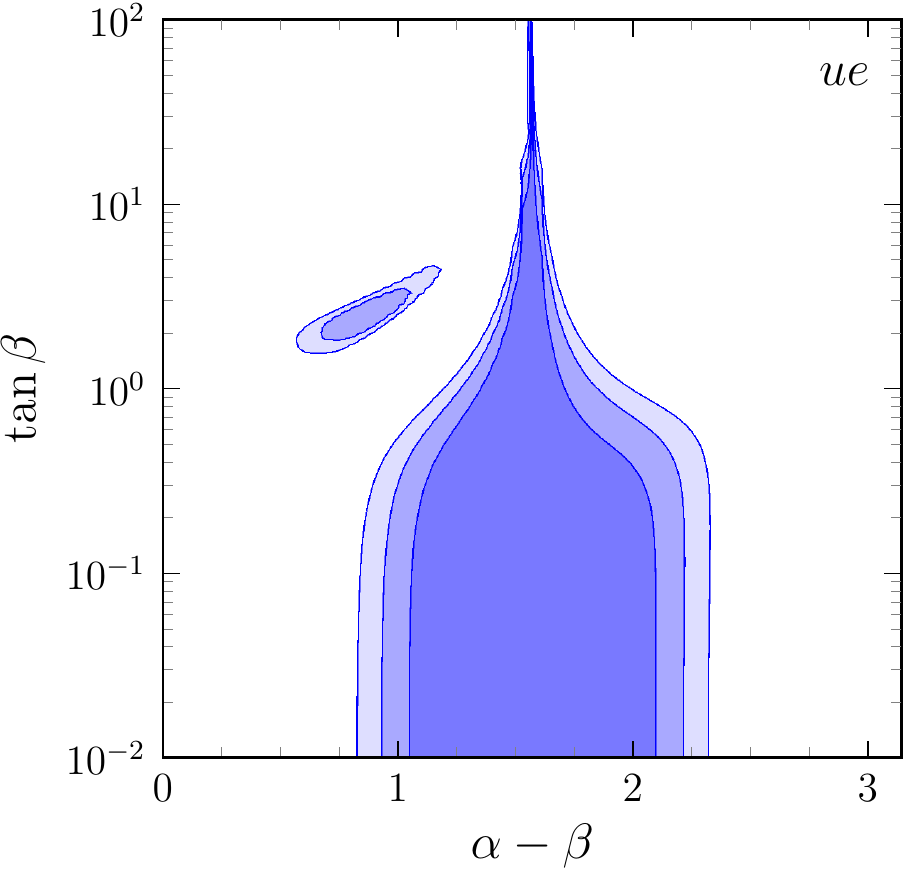}}\qquad
\subfigure[Model $b$ $\tau$]{\includegraphics[width=0.3\textwidth]{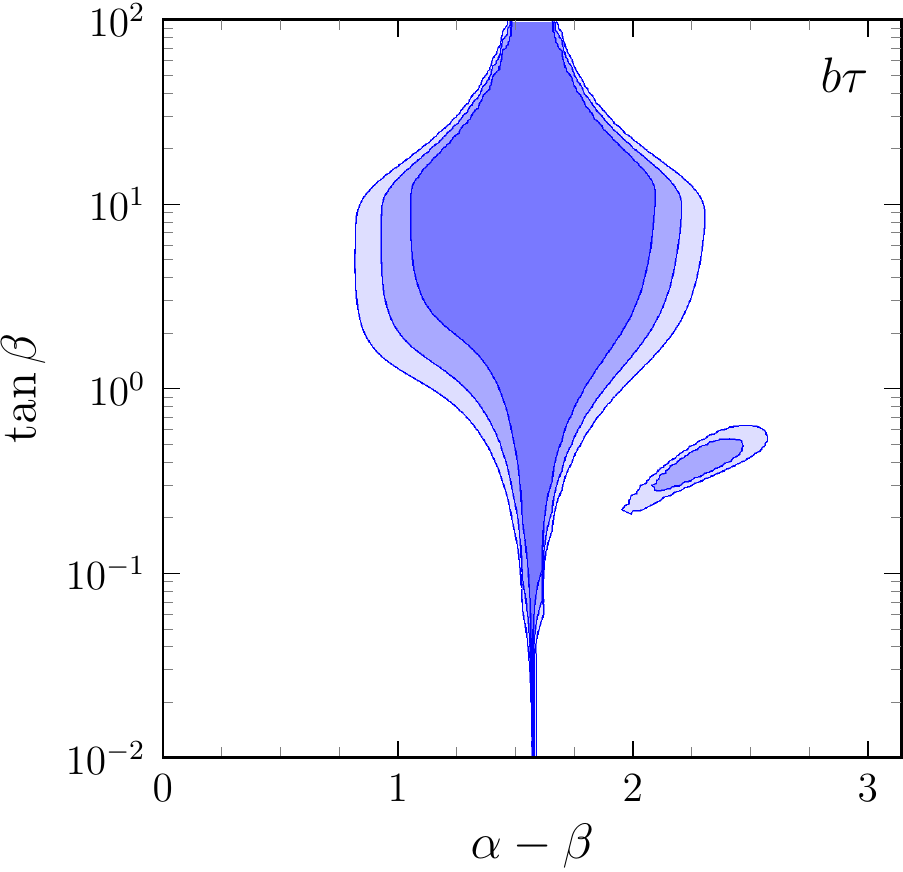}}
\caption{68\%, 95\% and 99\% CL regions in $\tb$ vs $\alpha-\beta$ for a small sample of models.\label{FIG:HiggsBase}}
\end{center}
\end{figure}

With all the ingredients in place, namely (i) the experimental constraints and (ii) the model predictions (simply expressed in terms of the different rescalings of SM couplings), a standard analysis of the $\{\tan\beta,\alpha-\beta\}$ parameter space can be built. As an illustration of the effect of imposing agreement with the set of constraints on flavour diagonal Higgs couplings, we show allowed regions in the $\log_{10}(\tan\beta)$ vs. $\alpha-\beta$ plane for a few models in Figure \ref{FIG:HiggsBase}. Since the overall agreement of different signal strengths with the SM is good, the region around $\cab=0$ is in all cases allowed. Depending then on the particular structure of the $\tan\beta$ dependences in the couplings, the $\alpha-\beta$ span of the allowed regions for large or small values of $\tan\beta$ can be anticipated. In addition it should be noticed that, in some cases, the fluctuations departing from signal strengths equal to 1, can be in fact accommodated with $\alpha-\beta\neq\pi/2$.

%%%%%%%%%%%%%%%%%%%%%%%%%%%%%%%%%%%%%%%%
\section{Constraints\label{AP:Constraints}}
%%%%%%%%%%%%%%%%%%%%%%%%%%%%%%%%%%%%%%%%
In this appendix we address the constraints imposed on $\ctti$ by (1) perturbativity requirements for the couplings in the scalar potential, (2) tree level contributions mediated by the three neutral scalars $\nh$, $\nH$, $\nA$, to the mixing amplitude $M_{12}$ in neutral meson systems and (3) two loop Barr-Zee contributions to the $\mu\to e\gamma$ decay branching ratio involving the flavour changing interactions of all three neutral scalars, impose on $\ctti$. It is important to stress that while $\nh\to\mu\tau$, $t\to \nh c,\nh u$ or $\nh\to bs, bd$, depend on $\ctti$ and no other unknown parameter related to the scalar sector, the constraints analysed in sections \ref{AP:Pert}, \ref{AP:Mix} and \ref{AP:MEG} of this appendix, do involve new parameters like the masses $m_\nH$ and $m_\nA$.
%%%%%%%%%%%%%%%%%%%%%%%%%%%%%%%%%%%%%%%%
\subsection{Perturbative unitarity\label{AP:Pert}}
%%%%%%%%%%%%%%%%%%%%%%%%%%%%%%%%%%%%%%%%
Neutral scalar masses and mixings arise from the scalar potential of the model and are related to the dimensionless quartic couplings $\lambda_i$. Perturbativity requirements like $\lambda_i\leq 4\pi$, could have some impact on the allowed values for $\ctti$. Following appendix D of \cite{Gunion:2002zf} (here $\lambda_5=\lambda_6=\lambda_7=0$, and furthermore notice in addition that in \cite{Gunion:2002zf} all $\lambda_i$ are two times our corresponding $\lambda_i$ in \refEQ{eq:ScalarPotential}),
\begin{equation}
m_A^2 \sab\cab=v^2[s_{2\alpha}(\lambda_2\sb^2-\lambda_1\cb^2)+(\lambda_{3}+\lambda_{4})s_{2\beta}c_{2\alpha})].\label{eq:pert:00}
\end{equation}
Since $\cab(\tb+\tbinv)=\frac{\cab}{\sb\cb}$, we have 
\begin{equation}
\frac{\cab\sab}{\sb\cb}=\cab(\tb+\tbinv)\sab=\frac{v^2}{m_{\nA}^2}[s_{2\alpha}(\lambda_2\tb-\lambda_1\tbinv)+2(\lambda_{3}+\lambda_{4})c_{2\alpha}].
\end{equation}
It is then clear that, for $m_\nA\sim v$, having $\ctti\sim\mathcal O(1)$ does not challenge naive perturbativity requirements like $\lambda_i\leq 4\pi$. For much larger values of $m_\nA$, however, the situation is more involved, and only a detailed analysis including all relevant parameters can gauge the precise extent of the constraints on $\ctti$ \emph{as a function of additional parameters}. This is beyond the scope of the present work.
Further relations, similar to \eqref{eq:pert:00}, but involving $m_\nH$ and $m_\nh$ instead of $m_\nA$, lead to the same conclusion on the perturbativity requirements for the $\lambda_i$'s versus the values of $\ctti$.\\
It is important to stress, however, that the presence of other constraints overrules the potential role of imposing perturbative unitarity in the scalar potential: $\nh\to\mu\tau$ alone, in neutrino type models, already requires $\ctti\lesssim 1$; bounds on rare top decays $t\to \nh c,\nh u$ in models of types $b$ and $s$ impose $\ctti<5$; for the remaining models constraints from meson mixings (addressed in the following), can play a more relevant role.
%%%%%%%%%%%%%%%%%%%%%%%%%%%%%%%%%%%%%%%%
\subsection{Mesons mixings\label{AP:Mix}}
%%%%%%%%%%%%%%%%%%%%%%%%%%%%%%%%%%%%%%%%
The contribution to the meson mixing amplitudes $M_{12}$ in BGL models is, to an excellent approximation (up to terms of order $\frac{m_d}{m_s}$, $\frac{m_d}{m_b}$, $\frac{m_s}{m_b}$ for up models, analogously up to $\frac{m_u}{m_c}$, $\frac{m_u}{m_t}$, $\frac{m_c}{m_t}$ terms for down models),
\begin{equation}
M_{12}\propto (\tb+\tbinv)^2\left(\frac{\cab^2}{m_h^2}+\frac{\sab^2}{m_H^2}-\frac{1}{m_A^2}\right).
\end{equation}
In the strict $\cab=0$ limit, for $m_A=m_H$ (a situation which does not clash with electroweak precision data), there is a complete cancellation at work. Departing from the $\cab= 0$ limit, in the three dimensional $\{\cab,m_\nH,m_\nA\}$ parameter space, there is now a two dimensional subspace where the cancellation is still complete
\[
(\cab^2,m_\nH^2,m_\nA^2)=(c^2,M^2,[c^2/m_\nh^2+(1-c^2)/M^2]^{-1}).
\]
Close to that subspace, considering the $\nh$ contribution alone, does not reflect the actual meson mixing constraint on $\cab$. With respect to that single $\nh$ contribution, a cancellation of one or even two orders of magnitude is achievable. However, larger cancellations that would be necessary in some cases to produce interesting phenomenological consequences (for example, for $(u,e)$ and $(u,\mu)$ models in figure \ref{FIG:hbdhbs:uplep:01}), may be less likely. Since a detailed analysis would involve additional parameters, $m_\nH$ and $m_\nA$, only would-be bounds from $\nh$ alone are explicitely shown in the different figures as an indication of potential constraint in some models. Other models, like all type $b$ models, are free from such constraints. 
% , (2) in order to avoid obscuring the interpretation that such plots in a full analysis, outside the scope of this work, would require.
%%%%%%%%%%%%%%%%%%%%%%%%%%%%%%%%%%%%%%%%
\subsection{$\mu\to e\gamma$\label{AP:MEG}}
%%%%%%%%%%%%%%%%%%%%%%%%%%%%%%%%%%%%%%%%

The radiative transition $\mu\to e\gamma$ in 2HDM is typically suppressed at the one loop level by lepton masses. Two loop contributions of the Barr-Zee type \cite{Barr:1990vd} where instead of two, only one suppressed scalar-fermion-antifermion coupling is involved -- thus reducing the chirality flip suppression -- can be dominant. The contribution from this class of diagrams in the context of 2HDM was addressed in \cite{Chang:1993kw}. In BGL models, this two loop prediction has the general structure
\begin{multline}
\text{Br}(\mu\to e\gamma)_{\rm 2\, loop}=\\
\frac{3}{8}\left(\frac{\alpha}{\pi}\right)^3(\tb+\tbinv)^2|U_{ej}U_{\mu j}^\ast|^2\,|A_{(Q)}|^2\simeq 5.77\times 10^{-9}\,(\tb+\tbinv)^2|U_{ej}U_{\mu j}^\ast|^2\,|A_{(Q)}|^2\,,
\end{multline}
where the neutrino-type model dependence is encoded in the prefactor $|U_{ej}U_{\mu j}^\ast|$ (which takes approximate values $0.32, 0.30, 0.11$ for $j=1,2,3$), and $A_{(Q)}$ is the amplitude
\begin{equation}
A_{(Q)}=\cab\sab\left[\Sigma(m_\nh)-\Sigma(m_\nH)\right]+\frac{8}{3}K_t\left[\cab^2f(z_\nh)+\sab^2f(z_\nH)-g(z_\nA)\right]\,,\label{eq:MUEG:01}
\end{equation}
with $y_X=M_W^2/M_X^2$ and $z_X=m_t^2/M_X^2$ ($X=\nh,\nH,\nA$), and $K_t$ the model dependent change in the coupling of scalars to the top quark in the loop in Table \ref{TAB:htt:MEG} \footnote{Notice that, owing to the hierarchy in the CKM matrix elements, as a very good approximation, in down models $d,s$, $K_t=\tb$, while in models $b$, $K_t=-\tbinv$.}. The functions $f$, $g$ and $h$ follow the definitions in \cite{Chang:1993kw}, and % $F_W(x)=3f(x)+5g(x)+3(g(x)+h(x))/4$.
\begin{equation}
\Sigma(m)=3f(y)+5g(y)+\frac{3}{4}\left[g(y)+h(y)\right]+\frac{f(z)-g(z)}{z}-\frac{8}{3}f(z)\,,\qquad y=\frac{M_W^2}{m^2}\,,\ z=\frac{m_t^2}{m^2}\,.
\end{equation}
\begin{table}[h!]
\begin{center}
\begin{tabular}{|c||c|c|c|}
\hline
Model type & $u,c$ & $t$ & Down $d_j$ \\\hline
Factor $K_t$ & $\tb$ & $-\tbinv$ & $(1-|\V{tj}|^2)\tb-|\V{tj}|^2\tbinv$\\\hline
\end{tabular}
\caption{$\nh t\bar t$ coupling, $K_t$ factor.}
\label{TAB:htt:MEG}
\end{center}
\end{table}

\begin{figure}[h!]
\begin{center}
\includegraphics[width=0.5\textwidth]{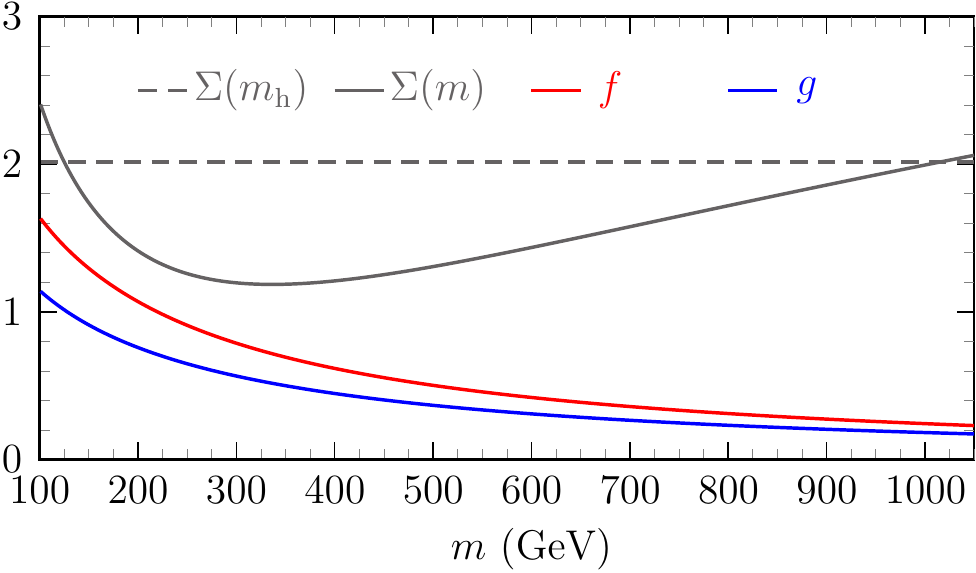}
\caption{$f(m_t^2/m^2)$, $g(m_t^2/m^2)$ and $\Sigma(m)$.\label{FIG:MEG:functions}}
\end{center}
\end{figure}

With the experimental bound \cite{Adam:2013mnn}, a simple estimate only considering the $\nh$-mediated contributions and $\Sigma(m_\nh)=2.016$ (neglecting the term $\cab^2f(z_\nh)$ for $\alpha-\beta\sim\pi/2$), puts the stringent bound $\cab(\tb+\tbinv)<0.031$. Nevertheless, ignoring the effect of the $\nH$ and $\nA$ terms is not justified. In Figure \ref{FIG:MEG:functions}, we plot $f(m_t^2/m^2)$, $g(m_t^2/m^2)$ and $\Sigma(m)$ as function of $m$, together with $\Sigma(m_\nh)$ as a reference \footnote{Although $\Sigma(m)$ would apparently induce non-decoupling contributions for large masses, $\Sigma(m)$ grows as $\ln m$ \cite{Chang:1993kw}, it appears multiplied by a $\cab$ factor and thus, taking into account the previous discussion on perturbativity, there is no ``real'' non-decoupling.}.
Attending to Eq.(\ref{eq:MUEG:01}) and Figure \ref{FIG:MEG:functions}, sizable cancellations among the different contributions can easily be at work. A simple exercise can be helpful to illustrate such scenarios: fixing $m_\nH=500$ GeV and requiring $\cab(\tb+\tbinv)>1$ (largely in excess of the bound that $\nh$ alone would give), one can numerically scan the range of values of $m_\nA$ for which the experimental bound is satisfied. For an initial scanned range $[100;1000]$ GeV, values of $m_\nA$ satisfying the requirements cover the full range. In other words, deriving bounds from $\nh$ alone is too simplistic. To remedy that and incorporate nevertheless the effect of $\mu\to e\gamma$, we follow two steps: (1) allowed $\tb$ vs $\alpha-\beta$ regions are first obtained from a scan of $M_\nH$ and $M_\nA$ values which fulfill the experimental bound on $\text{Br}(\mu\to e\gamma)$ for each model, (2) the general analysis of $\tb$ vs $\alpha-\beta$ in each model is then restricted to the previous regions. For the first step, two different regimes are explored, (a) $m_\nH$ and $m_\nA$ are independent, (b) $m_\nH$ and $m_\nA$ are required to be similar, namely $m_\nA\in[0.8;1.2]\times m_\nH$. Although the outcome of this auxiliary analysis is a reduction of the allowed $\tb$ vs. $\alpha-\beta$ regions, prospects for the different rare decays under consideration in $\nu$ models are not significantly altered.

\clearpage

\providecommand{\href}[2]{#2}\begingroup\raggedright\endgroup

\end{document}